\documentclass[11pt]{article}
\usepackage{amssymb,cite,epsf}
\newcommand\ttl[1]{`#1', }
\newcommand\toline[1]{--#1}
\newcommand{\phntpm}{{\phantom{\rlap{$\pm$}}}}
\newcommand{\phntn}{{\phantom{\rlap{(n)}}}}
\newcommand{\blank}[1]{}
\newcommand{\resection}[1]{\setcounter{equation}{0}\section{#1}}
\newcommand{\appsection}[1]{\addtocounter{section}{1} \setcounter{equation}{0}
                         \section*{Appendix \Alph{section}~~#1}}
\renewcommand{\theequation}{\thesection.\arabic{equation}}
\newcommand{\fract}[2]{{\textstyle\frac{#1}{#2}}}
\newcommand{\ri}{\right}

\newcommand{\lf}{\left}

\newcommand{\CS}{{\cal S}}

\newcommand\eq{\begin{equation}}
\newcommand\en{\end{equation}}
\newcommand\bea{\begin{eqnarray}}
\newcommand\eea{\end{eqnarray}}
\newcommand\nn{\nonumber}

\newcommand\Ai{{\rm Ai}}
\begin{document}
\begin{titlepage}
\vskip 0.5cm
\begin{flushright}
DTP-99/27 \\
ITFA 99-08 \\
{\tt hep-th/9906219} \\
June 1999
%%DRAFT: \today
\end{flushright}
\vskip 1.5cm
\begin{center}
{\Large{\bf On the relation between Stokes multipliers and the T-Q 
systems of conformal field theory
}}
\end{center}
\vskip 0.8cm
\centerline{Patrick Dorey%
\footnote{e-mail: {\tt p.e.dorey@durham.ac.uk}}
and Roberto Tateo%
\footnote{e-mail: {\tt tateo@wins.uva.nl}}}
\vskip 0.9cm
\centerline{${}^1$\sl\small Dept.~of Mathematical Sciences,
University of Durham, Durham DH1 3LE, UK\,}
\vskip 0.2cm
\centerline{${}^2$\sl\small Universiteit 
van Amsterdam, Inst.~voor Theoretische
Fysica, 1018 XE Amsterdam, NL\,}
\vskip 1.25cm
\begin{abstract}
\noindent
The vacuum expectation values of the
so-called ${\bf Q}$-operators of certain integrable quantum 
field theories have recently been identified with spectral determinants 
of particular Schr\"odinger operators. In this paper we extend the
correspondence to the ${\bf T}$-operators, finding that their vacuum
expectation values also have an interpretation as spectral determinants.
As byproducts we give a simple
proof of an earlier conjecture of ours, proved by another route by
Suzuki, and generalise a problem in ${\cal PT}$ symmetric quantum
mechanics studied by Bender and Boettcher.
We also stress that the mapping between ${\bf Q}$-operators 
and Schr\"odinger
equations means that certain problems in integrable quantum field
theory are related to the study of Regge poles in non-relativistic
potential scattering.
\end{abstract}
\end{titlepage}
\setcounter{footnote}{0}
\def\thefootnote{\fnsymbol{footnote}}
%%%%%%%%%%%%%%%%%%%%%%%%%%%%%%%%%%%%%%%%%%%%%%%%%%%%%%%%%%%%%%%%%%%%%
%%%  start of the paper %%%%%%%%%%%%%%%%%%%%%%%%%%%%%%%%%%%%%%%%%%%%%
%%%%%%%%%%%%%%%%%%%%%%%%%%%%%%%%%%%%%%%%%%%%%%%%%%%%%%%%%%%%%%%%%%%%%
%
\resection{Introduction and review}
\label{introsec}
An unexpectedly precise connection between certain two-dimensional
quantum field theories and the quantum mechanics of anharmonic
oscillators
was uncovered in~\cite{DT3}.
The quantum-mechanical side of this correspondence concerned the
Schr\"odinger problem
\eq
\hat{H} \psi(x)= \lf ( - {d^2 \over dx^{2}} + |x|^{2M} \ri )
\psi(x) = E\psi(x)
\label{Sch}
\en
on the real line $x\in(-\infty,\infty)\,$,
and the associated spectral determinant $D_M(\lambda)$, defined
for $M>1$ as
\eq
D_{M}(E)=D_{M}(0)\prod_{k=0}^{\infty}\lf(1-{E\over E_k}\ri),
\en
with the product running over the set $\{E_k\}$ of eigenvalues of
$\hat{H}$.
The normalisation $D_M(0)=\sin(\pi/(2M{+}2))^{-1}$ is automatic if
$D_M$ is defined as a zeta-regularized functional
determinant (see~\cite{V1}), and is also the natural choice for
current purposes. However we have chosen to negate
$E$ as compared to the conventions of \cite{DT3,V1}, so that the
zeroes of $D_M(E)$ as defined here
coincide with the eigenvalues of $\hat H$ rather
than their negatives.
 
The first result of \cite{DT3} was a mapping between the functional
relation
satisfied by $D_2(E)$, discovered in~\cite{V0},
and another set of functional relations,
known as a Y-system, which had previously arisen in the study of
perturbed
conformal field theories~\cite{Zam2}. (The precise relation given
in \cite{V0} had in fact also cropped up in the context of
integrable lattice models, in~\cite{Pearce}.)
One consequence was a novel
expression for $D_2(E)$
as the solution of a certain non-linear integral equation,
a result that was fixed uniquely
by the functional relations and known
analyticity properties on
the two sides of the correspondence.
The move to higher integer values of $M$ brings a
considerable increase in the complexity of the functional relations.
In~\cite{DT3}
the spectral problems for $M>2$ were related to certain further
sets of functional relations, associated with the Y-systems,
called T-systems, but this result
rested on a combination of analytical and numerical evidence
without being
proved. This gap was filled by Suzuki in~\cite{Suz1};
an alternative proof will be
given below.
 
The other strand in~\cite{DT3} was a relation between
the even and odd spectral
subdeterminants for the problem~(\ref{Sch}) and the vacuum
expectation values of objects
known as ${\bf Q}$-operators~\cite{Bax1,BLZ2}, taken at certain
values of the `Virasoro parameter' $p$.
The subdeterminants
are defined via a split of the eigenvalues according
to the parity of their eigenfunctions, as
\eq
D_M^{\pm}(E)
= D_M^{\pm}(0)\prod_{k {{\rm even}\atop{\rm odd}}}
\lf(1-{E \over E_k}\ri)
\en
(so that $D_M(E)=D_M^+(E)D_M^-(E)\,$), while
${\bf Q}$-operators ${\bf Q}_+$ and ${\bf Q}_-$
were introduced, in the form that we shall need, in~\cite{BLZ2}. 
They are continuum analogues of the
Q-matrices of Baxter~\cite{Bax1}. To state the result of~\cite{DT3}
more precisely, we recall that in the construction of \cite{BLZ2},
the operators ${\bf Q}_{\pm}$ act on Virasoro modules labeled by
$p$, with central charge $c=1-6(\beta{-}\beta^{-1})^2$, $\beta$
being a further parameter. In each module the `vacuum'
(highest weight) state
$|p\rangle$ has conformal dimension $(p/\beta)^2+(c{-}1)/24$,
and is an eigenstate of the ${\bf Q}$-operators. Setting
\eq
A_{\pm}(\lambda,p)=\lambda^{\mp 2p/\beta^2}
\langle p|{\bf Q}_{\pm}(\lambda)|p\rangle~,
\label{adef}
\en
the eigenvalues
$A_{\pm}(\lambda,p)$ are entire functions of the square of the
spectral parameter $\lambda$, and we have
\eq
A_{\pm}(\lambda,\beta^2/4)=
\alpha^{\mp}D^{\mp}_M(\lambda^2/\nu^2)
\label{AaD}
\en
where $M=\beta^{-2}{-}1$,
$\nu=(\frac{1}{2}\beta^2)^{1-\beta^2}\Gamma(1{-}\beta^2)^{-1}$, and
the proportionality constants $\alpha^{\pm}$ are rederived
below.
 
In contrast to the first result, $M$ does not have to be an integer
for this to hold, and so the modulus signs in
(\ref{Sch}) are in general obligatory when the problem is posed on the
full real line. However (as exploited in \cite{V4} for $M$
half-integral) we can alternatively set up the problem on the
half-line $(0,\infty)$ with the potential $x^{2M}$, so long as
the boundary condition at $x=0$ is chosen correctly. The even
wavefunctions are picked out by the Neumann condition
$\psi'_{2k}(0)=0$, and the odd by the Dirichlet condition
$\psi_{2k+1}(0)=0$. These two spectral problems thus yield the even
and odd spectral subdeterminants for the whole-line problem
directly.
 
The problem of finding spectral determinants related to the
${\bf Q}$-operators at general values of $p$ was addressed
in~\cite{BLZschr}, where it was found that the problem~(\ref{Sch})
should be modified to
\eq
\lf( - {d^2 \over dx^{2}} + x^{2M}
+\frac{l(l{+}1)}{x^2} \ri)
\psi(x) = E\psi(x)
\label{Schl}
\en
on the half-line $(0,\infty)$. Imposing the two
possible power-like behaviours
at $x=0$, namely $\psi(x)\propto x^{l+1}$ and $\psi(x)\propto
x^{-l}$, results in two different spectral problems, and their
spectral determinants are proportional to the functions
$A_+(\lambda,(2l{+}1)\beta^2/4)$ and
$A_-(\lambda,(2l{+}1)\beta^2/4)$
respectively.
 
Much of this work rested on the so-called quantum Wronskian relation
satisfied by the operators ${\bf Q}_+$ and ${\bf Q}_-$~\cite{BLZ2}.
But a key feature of~\cite{Bax1,BLZ2} was another
functional equation, relating the ${\bf Q}$-operators
to further operators ${\bf T}(\lambda)$, sometimes called quantum
transfer matrices. Called the
T-Q relation, it is
\eq
{\bf T}(\lambda){\bf  Q}_{\pm}(\lambda)=
{\bf Q}_{\pm}(q^{-1}\lambda)+{\bf Q}_{\pm}(q\lambda)\,,
\en
where
\eq
q=e^{i\pi\beta^2}.
\label{omegabeta}
\en
The vacua $|p\rangle$ are also eigenstates
of the ${\bf T}(\lambda)$, and
if we set $T(\lambda,p)=\langle p|{\bf T}(\lambda)|p\rangle$ then the
T-Q relation for these vacuum eigenvalues can be written as
\eq
T(\lambda,p)A_{\pm}(\lambda,p)= 
e^{\mp 2\pi ip}A_{\pm}(q^{-1}\lambda,p)
+ e^{\pm 2\pi ip}A_{\pm}(q\lambda,p)\,,
\label{tqeq}
\en
an equation that was also obtained in~\cite{FLSa}.

In this paper we point out that this relation
also has a natural interpretation in the context of the
Schr\"odinger equation, thus finding a r\^ole for the ${\bf T}$
operators at general $\beta$ and $p$
in the `Schr\"odinger picture'.
This leads to an alternative derivation of the results of
\cite{DT3,BLZschr}, and a novel interpretation of the fusion relations 
and their truncations. It also has the bonus, for 
integer values of $M$, of providing a
simple proof of the T-system conjecture alluded to above.
All of this material is contained in \S\S \ref{tqsec}--\ref{yssec},
while \S\ref{bbsec} applies the results to a problem in quantum
mechanics, and \S\ref{dualsec} discusses duality properties. Finally,
\S\ref{concsec} contains our conclusions, and an appendix details the
calculation of a certain asymptotic used in \S\ref{tqsec}.

\resection{The general T-Q relation}
\label{tqsec}
The results of~\cite{Si} provide a convenient framework for
our discussion, and we begin with a quick
summary of some of this material.
 
Consider the differential equation
\eq
\lf(-\frac{d^2}{dx^2}+P(x)\ri)\psi(x)=0
\label{sibeq}
\en
for general complex values of $x$. For our purposes we must set
$P(x)=x^{2M}-E+l(l{+}1)x^{-2}$.  In \cite{Si}
$P(x)$ was taken to be a polynomial in $x$, restricting
$2M$ to be
a positive integer, and $l(l{+}1)$ to be zero. 
Allowing $2M$ to be a general real number larger than $-2$, or
$l(l{+}1)$
to be nonzero, should not change the result quoted below in any essential
way, though it will usually introduce a branch point in
$\psi(x)$ at $x=0$. 
Modulo this proviso, we have \cite{Si}:\\
$\bullet$ The equation~(\ref{sibeq}) has a solution $y=
y(x,E,l)$
such that
 
\noindent
(i) $y$ is an entire function of $(x,E)$ (though, for the reason just 
mentioned, $x$ must in general be considered to live on a 
suitable cover of the punctured complex plane).
 
\noindent
(ii) $y$ and $y'= dy/dx$ admit the asymptotic representations
\eq
y\sim x^{-M/2}
\,\exp(-\fract{1}{M+1}x^{M+1})
\label{asrep}
\en
\eq
\! y'\sim -x^{M/2}
\,\exp(-\fract{1}{M+1}x^{M+1})
\label{dasrep}
\en
as $x$ tends to infinity in any closed sector contained in the
sector
\eq
|\arg x\,|<\frac{3\pi}{2M+2}~.
\en
\noindent
Furthermore, the solution $y$ is uniquely characterised by this
information. (Note though
that the asymptotic (\ref{asrep}) is only valid as given
if $M>1$. Extra terms
appear for $M\le 1$, consistent with the WKB
result that the solution which decays as $x\rightarrow+\infty$
is asymptotically proportional
to $P(x)^{-1/4}\exp(-\int^x P(t)^{1/2}dt)\,$.)
 
Let $\CS_k$ denote the sector
\eq
\lf|\arg x - \frac{2k\pi}{2M{+}2}\ri|<\frac{\pi}{2M{+}2}\,.
\en
{}From (\ref{asrep}) it follows that $y$ tends
to zero as $x$ tends to infinity in $\CS_0$, and to infinity as
$x$ tends to infinity in $\CS_{-1}$ and $\CS_1$. More technically,
one says that $y$ is subdominant in $\CS_0$, and dominant in
$\CS_{\pm 1}$. To find solutions subdominant in other sectors,
consider $\hat{y}(x)=
y(ax,E,l)$ for any constant $a$. This function solves
the equation
\eq
\lf(-\frac{d^2}{dx^2}+a^{2M+2}x^{2M}-a^2E+
\frac{l(l{+}1)}{x^{2}}\ri)\hat{y}(x)=0\,.
\en
Thus if $a^{2M+2}=1$,  $y(a x,a^{-2}E,l)$ is another solution
to the original problem (\ref{sibeq}). Setting $\omega=
\exp(\pi i/(M{+}1))$ 
we therefore have a set of solutions
\eq
y_k\equiv y_{k}(x,E,l)=
\omega^{k/2}y(\omega^{-k}x,\omega^{2k}E,l)\,,
\label{ykdef}
\en
with $y_k$ subdominant in $\CS_k$
and dominant in $\CS_{k\pm 1}$. (Our convention
differs from \cite{Si} by the factor of $\omega^{k/2}$,
which is included for later convenience.)

A consequence of these facts is that
each pair $\{y_k,y_{k+1}\}$ provides a set of linearly independent
solutions to the second-order equation (\ref{sibeq}), and any
other solution can be expanded in terms of them. In particular,
\eq
y_{k-1}(x,E,l)=C_k(E,l)y_{k}(x,E,l)+\tilde C_k(E,l)y_{k+1}(x,E,l)\,.
\label{stokedef}
\en
The functions $C_k$ and $\tilde C_k$ are called the Stokes
multipliers for $y_{k-1}$ with respect to $y_{k}$ and $y_{k+1}$.
It follows from (\ref{ykdef}) that $C_k(E,l)=C_{k-1}(\omega^{2}E,l)$ and
$\tilde C_k(E,l)=\tilde C_{k-1}(\omega^{2}E,l)$. For brevity we will
write $C_{0}$ and $\tilde C_{0}$ as $C$ and $\tilde C$ respectively.
(Again, we differ slightly from the conventions of \cite{Si}, where the
abbreviations $C$ and $\tilde C$ were instead reserved for 
$C_1$ and $\tilde C_1$.)

The Stokes multipliers can be expressed in terms of
Wronskians. Recall that the Wronskian $W[f,g]$ of two functions
$f(x)$ and $g(x)$ is defined as
\eq
W[f,g]=fg'-f'g\,.
\en
If $f$ and $g$ both solve a Schr\"odinger equation such as
(\ref{sibeq}), then $W[f,g]$ is independent of $x$; furthermore, it
vanishes if and only if $f$ and $g$ are linearly dependent.
Taking the Wronskian of (\ref{stokedef}) at $k=0$ with $y_1$
and $y_0$ shows that
\eq
C=\frac{W_{-1,1}}{W_{0,1}}\quad,\quad
\tilde C=-\frac{W_{-1,0}}{W_{0,1}}\,,
\label{cwronk}
\en
where we used the abbreviation $W_{k_1,k_2}$ for $W[y_{k_1},y_{k_2}]$.
These Wronskians are entire functions of $E$ and $l$. Since $y_0$
and $y_1$ are independent, $W_{0,1}$ never vanishes, and
$C$ and $\tilde C$ are also entire.

In fact, all of the $\tilde C_k$ are identically equal to $-1$~\cite{Si}.
This follows
from (\ref{cwronk}) and
the relations $W_{k_1+1,k_2+1}(E,l)=
W_{k_1,k_2}(\omega^{2}E,l)$ and
$W_{0,1}(E,l)=2i$. 
(The second of these is found by
evaluating $W_{0,1}$ as $x$ tends
to infinity in the sectors $\CS_0$ or $\CS_1$, where the
asymptotic behaviours of $y_0$ and $y_1$ and their derivatives
are determined by (\ref{asrep}) and (\ref{dasrep}).)
Since $y_{-1}(x,E,l)=y_1(x^*,E^*,l^*)^*$, 
it also follows from (\ref{cwronk})
that $C(E,l)$ is real whenever $E$ and $l$ are real.

The basic Stokes relation (\ref{stokedef}) at $k=0$ is therefore
\eq
C(E,l)y_0(x,E,l)=y_{-1}(x,E,l)+y_1(x,E,l)
\label{tqyrel}
\en
with
\eq
C(E,l)=\frac{1}{2i}W_{-1,1}(E,l)\,.
\label{cexp}
\en
If (\ref{tqyrel}) is rewritten
in terms of $y$ it becomes
\eq
C(E,l)y(x,E,l)=
\omega^{-1/2}y(\omega x,\omega^{-2}E,l)+
\omega^{1/2}y(\omega^{-1}x,\omega^2E,l)
\label{tqyrelx}
\en
With $x$ formally set to zero, this
has exactly the form of (\ref{tqeq}) for
$A_+(\lambda,p)$,
albeit at the specific value $\beta^2/4$ of $p$ for which
$e^{2\pi ip}=\omega^{1/2}$. (It also matches the
T-Q relation for $A_-$ at $p=-\beta^2/4$, but since $A_-(\lambda,p)
=A_+(\lambda,-p)$~\cite{BLZ2} this is not an independent result.)
However this tactic
only works when $l(l{+}1)=0$. 
Otherwise, the resulting equation in $E$ is either trivial or
meaningless: if $l(l{+}1)<0$, that is $-1<l<0$,
then $y(x{=}0,E,l)$ is identically
zero, while if $l(l{+}1)>0$ then $y(x{=}0,E,l)$ is almost everywhere
infinite.
 
The problem arises because any solution to (\ref{sibeq}) 
is a linear combination of one solution, $\psi^+$,
behaving near $x=0$ as $x^{l+1}$, and
one, $\psi^-$, behaving there as $x^{-l}$. Both
of these are zero at $x=0$ if $l(l{+}1)<0$, whilst one or other
is infinite if $l(l{+}1)>0$. However,
rather than considering the functions $y(x,E,l)$ at $x{=}0$ directly, 
we can take a hint from the result of \cite{BLZschr} and
project onto either $\psi^+$ or $\psi^-$.
We choose to fix $\psi^+$ 
by the $x\rightarrow 0$ asymptotic
\eq
\psi^+(x,E,l)\sim x^{l+1} + O(x^{l+3})\,.
\label{psidef}
\en
Since $\psi^-$, the other solution,
behaves as $x^{-l}$, this only determines
$\psi^+$ uniquely if $\Re e\, l>-3/2$. If
necessary, $\psi^+$ can be defined
outside this domain by analytic continuation. In particular, since
$l$ only appears in (\ref{sibeq}) in the combination
$l(l{+}1)$\,, we can continue from $l$ to $-1{-}l$ and define
$\psi^-$ as
\eq
\psi^-(x,E,l)\equiv \psi^+(x,E,-1{-}l)\,.
\label{psidefmin}
\en
This procedure
does bring some subtleties, to which we shall return 
\S\ref{tqgenrel} below, but they do not affect the arguments of
this section. In discussions of the radial
Schr\"odinger equation (see, for example, chapter 2 of \cite{Euan} or
chapter 4 of \cite{NEWT}\,) 
$\psi^+(x,E,l)$ for $\Re e\, l>-1/2$ is sometimes
called the regular solution.

In analogy to (\ref{ykdef}),
we define `shifted' solutions
$\psi^{\pm}_k\,$:
\eq
\psi^{\pm}_k\equiv \psi^{\pm}_k(x,E,l)
=\omega^{k/2}\psi^{\pm}(\omega^{-k}x,\omega^{2k}E,l)\,.
\label{shifted}
\en
These also solve the original problem (\ref{sibeq}). By
considering the $x\rightarrow 0$ limit it is easily
seen that
\eq
\psi^{\pm}_k(x,E,l)=\omega^{\mp k(l+1/2)}\psi^{\pm}(x,E,l)\,.
\en
We also have
$W[y^{\phntpm}_{k_1+1},\psi^{\pm}_{k_2+1}](E,l)=
W[y^{\phntpm}_{k_1},\psi^{\pm}_{k_2}](\omega^2E,l)\,$,
so
\bea
W[y^{\phntpm}_{k},\psi^{\pm}](E,l) &=&
\omega^{\pm k(l+1/2)}\,W[y^{\phntpm}_{k},\psi^{\pm}_k](E,l) \nn\\[3pt]
&=&
\omega^{\pm k(l+1/2)}\,W[y,\psi^{\pm}](\omega^{2k}E,l)\,.
\label{wres}
\eea
We can now take the Wronskian of both sides
of (\ref{tqyrel}) with $\psi^{\pm}$. Defining
\eq
D^{\mp}(E,l)\equiv W[y(x,E,l),\psi^{\pm}(x,E,l)]
\label{ddef}
\en
and using equation (\ref{wres}),
the Stokes relation (\ref{tqyrel}) becomes
\eq
C(E,l)D^{\mp}(E,l)=
\omega^{\mp(1/2+l)}D^{\mp}(\omega^{-2}E,l)+
\omega^{\pm(1/2+l)}D^{\mp}(\omega^2E,l)
\label{cdeq}
\en
and (\ref{tqeq}) has indeed been matched, provided
$\omega$ is equal $q$
(that is, $e^{i\pi/(M{+}1)}=e^{i\pi\beta^2}$), and
$\omega^{l+1/2}$ is equal to
$e^{2\pi ip}$.
These are the same relations between $M$ and
$\beta$, and $l$ and $p$, as obtained 
in \cite{DT3,BLZschr}, found here by an alternative route. 

To establish the precise relation between the functions appearing 
in equations
(\ref{tqeq}) and (\ref{cdeq}), we can use the fact that, when
combined with certain analyticity properties, T-Q relations of
this kind are extremely restrictive~\cite{Bax1,KBP,BLZ2}. 
Since $D^+(E,l)=D^-(E,-l{-}1)$, we need only
consider 
$D^-(E,l)$.
In addition to (\ref{cdeq}), we have
\begin{itemize}
\item[(i)] $C$ and $D^-$ are entire functions of $E\,$;
\item[(ii)] If $l$ is real and larger than $-1/2$, then 
the zeroes of $D^-$ all lie on the positive real axis of
the complex-$E$ plane;
\item[(iii)] If $-1{-}M/2<l<M/2$, then
the zeroes of $C$ all lie away from the positive real axis of
the complex-$E$ plane;
\item[(iv)] If $M>1$ then $D^-$ has the large-$E$ asymptotic
\eq
\log D^-(E,l)\sim { a_0 \over 2}(-E)^{\mu}~~,\qquad 
|E|\rightarrow \infty~,~~|\arg(-E)\,|<\pi
\label{Dlim}
\en
where $\mu=(M{+}1)/2M$ and
\eq
a_0= 2 \int_0^{\infty}[(t^{2M}+1)^{1/2}-t^M]dt
=-\frac{1}{\sqrt\pi}
\Gamma(-\fract{1}{2}-\fract{1}{2M})\Gamma(1+\fract{1}{2M})\,;
\label{intresult}
\en
\item[(v)] If $E=0$ then\\[-12pt]
\eq
D^-(0,l)=
\frac{1}{\sqrt{\pi}} 
\Gamma(1+\fract{2l+1}{2M+2})\,
(2M{+}2)_{\phantom{+}}^{\frac{2l+1}{2M+2}+ \frac{1}{2}}\,.
\label{norm}
\en
\end{itemize}
Property (i) follows from the definition (\ref{ddef}) of $D^-$ as a
Wronskian, given that the functions involved are themselves entire functions 
of $E$. Property (ii) is also straightforward, since a zero of $D^-(E,l)$
signals the existence of an eigenfunction for 
(\ref{Schl})
at that value of $E$, decaying as $x^{l+1}$ as $x\rightarrow 0$,
and exponentially as $x\rightarrow +\infty$. 
The self-adjoint
nature of this problem for $l>-1/2$ then ensures the reality of
these zeroes. For $l>0$, the potential is everywhere positive and
multiplying (\ref{Schl}) by $\psi^*(x)$ and
integrating from $0$ to $\infty$ shows that all of
the eigenvalues $E$ must also all be positive. For $-1/2<l<0$ the
centrifugal term in the potential, $l(l{+}1)/x^2$, is negative but the
same style of argument can be applied to the transformed equation 
(\ref{tranfeq}), with the conclusion that the eigenvalues are again all 
positive.
Property (iii) is more delicate,
and 
further
discussion will be postponed until \S\ref{shosec},
where a partial result will be established.
Finally, property (iv) follows from a WKB analysis, which is
outlined
in appendix A, and property (v) from a mapping of the problem at
$E=0$ into an
exactly-solvable case,
given in \S\ref{shosec} below.

We now claim that for $M>1$ and $-1/2<l<M/2\,$,
the T-Q relation (\ref{cdeq}) and
properties (i)--(v) characterise the functions $C(E,l)$ 
and $D^-(E,l)$ uniquely. 
Furthermore, with the identifications
\eq
\beta^2=\frac{1}{M{+}1}\quad,\quad p=\frac{2l{+}1}{4M{+}4}
\en
the same T-Q relation and the same properties (i)--(v)
hold for the functions
$T(\lambda,p)$ and $A_+(\lambda,p)$ of~\cite{BLZ2},
save for $\lambda^2$ replacing $E$ in (i),
the asymptotic in (iv) becoming
\eq
\log A_{+}(\lambda,p)\sim
(M{+}1)\Gamma\bigl(\fract{1}{2\mu}\bigr)^{2\mu}%
a_0\,(-\lambda^2)^{\mu}\,,\qquad 
|\lambda^2|\rightarrow \infty~,~~|\arg(-\lambda^2)\,|<\pi\,,
\!\!\!\!\!\!\!\!\!
\en
and $A_+(0,p)$ being equal to one rather than the value given in (v).
The asymptotics can be made to agree by setting
\eq
\lambda=\nu E^{1/2}~,\qquad 
\nu=(2M{+}2)^{-1/2\mu}\Gamma(\fract{1}{2\mu})^{-1}
\en
while (v) is fixed by multiplying $A_+(\lambda,p)$ by $D^-(0,l)$. 
By the claimed uniqueness, we therefore have
\eq
A_{\pm}(\lambda,p)\Bigl|_{\beta^2}{~}={~}
\alpha^{\mp}D^{\mp}\!\lf(\lf(\fract{\lambda}{\nu}\ri)^{2\!\!},%
\fract{2p}{\beta^{2}}{-}\fract{1}{2}\ri)\Bigl|_{M=\beta^{-2}-1}~,
\label{adrel}
\en
where $\alpha^{\mp}=D^{\mp}(0,2p/\beta^{2}{-}1/2)^{-1}$,
and
\bea
\quad~{}
T(\lambda,p)\Bigl|_{\beta^2}&=&
C\!\lf(\lf(\fract{\lambda}{\nu}\ri)^{2\!\!},%
\fract{2p}{\beta^{2}}{-}\fract{1}{2}\ri)\Bigl|_{M=\beta^{-2}-1}\nn\\[3pt]
&=&
\frac{1}{2i}W_{-1,1}\!\lf(\lf(\fract{\lambda}{\nu}\ri)^{2\!\!},%
\fract{2p}{\beta^{2}}{-}\fract{1}{2}\ri)\Bigl|_{M=\beta^{-2}-1}~.
\label{twrel}
\eea
Continuation from $l$ to $-l{-}1$ and the identity
$A_-(\lambda,p)=A_+(\lambda,-p)$ were used to deduce the relation
between $A_-$ and $D^+$ in (\ref{adrel}), while
the result (\ref{cexp}) was used to express $C(E,l)$ in terms of
$W_{-1,1}(E,l)\,$ in (\ref{twrel}). As will seen in the next section, 
the expression in terms
of the Wronskian appears to be the most general.

It remains to discuss the uniqueness property.
Observe first that the asymptotic (\ref{Dlim}) applies
as $|E|\rightarrow\infty$ along any ray with $\arg E\neq 0$, since the 
WKB problem only has turning-points if $E$ is on the positive real axis.
Since the growth of $\log D^-(E,l)$ in this remaining direction is no 
greater we conclude 
that the order of $D^-(E,l)$ as a function of $E$
is equal to $(M{+}1)/2M$. (The order of an entire
function $f(z)$, if finite, is the lower bound of all positive numbers $A$
such that $|f(z)|=O(e^{r^A})$ as $|z|=r\rightarrow\infty$. See, for example,
chapter 8 of
\cite{titch}.) For $M>1$, $(M{+}1)/2M$ is less than~$1$, and Hadamard's
factorization theorem implies that $D^-$ can be written as
an infinite product over its set of zeroes $\{E_n\}$ of the
following form:
\eq
D^-(E,l)=D^-(0,l)\prod_{n=0}^{\infty}\lf(1-\frac{E}{E_n}\ri)\,.
\label{factor}
\en
As an aside, we note that this is the final ingredient 
needed to prove that $D^-$, and hence $A_+$,
is proportional to a spectral
determinant: from the discussion of property (ii) above, the zeroes of
$D^-$ coincide with those of the relevant spectral determinant, and
the residual ambiguity of an entire function with no zeroes is reduced
to an overall constant
by (\ref{factor}). (In fact, the normalisation of $D^-$ given by
(\ref{norm}) coincides with an earlier
`natural' normalisation for the spectral determinant
when $M$ is an integer and $l$ is equal to $0$ or $1$ \cite{V4}, so it
is reasonable to say that $D^-$ is actually equal to the spectral
determinant.)

If $M\leq 1$ the order is larger, 
permitting a more complicated
prefactor to the product in (\ref{factor}), and furthermore the asymptotic
density of zeroes necessitates modifications
to the terms in the
product to ensure their convergence (again, see~\cite{titch}).
For the most part we 
have been excluding such cases from our discussion, but we
remark that there is evidence that many of our results continue
to hold even after the region $M>1$ (called the `semiclassical domain' in
\cite{BLZ2}) is left. Some of this evidence was given in~\cite{DT3};
some more can be found in the next section, and in 
\S\ref{bbsec}.

Reverting to $M>1$, the product representation (\ref{factor}) 
allows the steps
described in \cite{DDV,BLZ2} to be repeated to show that $D^-$ is determined
by a nonlinear integral equation of the type introduced 
in~\cite{KBP,DDV}. Introduce the function
\eq
d(E,l)=\omega^{2l+1}\frac{D^-(\omega^2E,l)}{D^-(\omega^{-2}E,l)}\,.
\en
{}From (\ref{cdeq}) and property (i), the points at which $d=-1$ are
exactly the
zeroes of $C$ and $D^-$. By properties (ii) and (iii), those on the
positive real axis are the zeroes of $D^-$, those on the negative
real axis the zeroes of $C$. (The need for property
(iii) was not mentioned in \cite{BLZ2}, but if it fails equation
(\ref{nlie}) below
has to be modified, becoming the massless
version of one of the equations obtained in \cite{FMQR}.)
The large-$E$ behaviour of $d(E,l)$ follows from 
(\ref{Dlim})\footnote{this asymptotic is only correct for $M>1$. The
situation for $M<1$ is discussed in appendix~A}:
\eq
\log d(E,l)\sim \left\{
\begin{array}{ll}
-\fract{1}{2}ib_0(1{-}e^{-i\pi/M}) (E)^{\mu} \quad
 & ~\frac{2\pi}{M{+}1}<\arg (E)<2\pi-\frac{2\pi}{M{+}1}\\[7pt]
-\fract{1}{2}ib_0 (E)^{\mu} 
 & -\frac{2\pi}{M{+}1}<\arg (E)<\frac{2\pi}{M{+}1}\\[7pt]
-\fract{1}{2}ib_0(1{-}e^{i\pi/M}) (E)^{\mu} 
 & -2\pi+\frac{2\pi}{M{+}1}<\arg (E)<-\frac{2\pi}{M{+}1}
\end{array}\right.
\en
where $b_0=2\cos(\frac{\pi}{2M})a_0$.
(By $(E)^{\mu}$ we imply $e^{i\mu\arg(E)}|E|^{\mu}$. Thus the
first and third asymptotics coincide, as indeed
they must
since $d$ is a single-valued function of $E$. In the language of
\cite{DDV}, they correspond to the `second determination'.)

Now trade $E$ for the new variable
$\theta=\mu\log(\nu^2E)$.
Then the arguments of \cite{BLZ2} can be followed to show that
the function $f(\theta,l)=\log
d(\nu^{-2\!}e^{\theta/\mu\!},l)$ solves
\bea
f(\theta,l)&=&
i \pi ( l{+}\fract{1}{2})  -\fract{1}{2}i\,b_0\nu^{-2\mu}e^{\theta}
+\int_{{\cal
C}_1}\!\varphi(\theta{-}\theta')\log(1{+}e^{f(\theta'\!,l)})\,d\theta'
\nn\\[2pt]
&&\qquad\qquad\qquad\qquad\qquad{}-
\int_{{\cal
C}_2}\!\varphi(\theta{-}\theta')\log(1{+}e^{-f(\theta'\!,l)})\,d\theta'
\label{nlie}
\eea
where the contours ${\cal C}_1$
and ${\cal C}_2$ run from $-\infty$ to $+\infty$, just below and just
above the real $\theta$-axis, and
\eq
\varphi(\theta)=\int_{-\infty}^{\infty}%
\frac{e^{i k\theta}\sinh\fract{\pi}{2}(\xi{-}1) k}%
{2\cosh\fract{\pi}{2} k \sinh\fract{\pi}{2}\xi k }%
\frac{d k}{2\pi}~~~,\qquad \xi=\fract{1}{M}\,.
\label{krnl}
\en
As mentioned above, such nonlinear integral equations first arose 
in~\cite{KBP,DDV}.
We must now assume that the solution to this equation is unique. In
this (`massless') context such an assumption seems reasonable, though
we did not attempt a complete proof. (Note though that for
the closely-related TBA integral equations, such a uniqueness 
result can fail at nonzero complex values of the mass scale, 
as can be seen
from figure~1 of \cite{DT1}.)
A knowledge of $f$ fixes $D^-$ up to an overall constant, given
the general information on the locations of the zeroes of $D^-$ and
$C$ contained in properties (i) and (ii) above. In turn, this
constant is determined by property (v), and so $D^-$ and hence $C$ 
have been determined.
While a more direct approach to the main uniqueness
claim would perhaps be more satisfactory, the fact, already stressed
in \cite{DT3},  that spectral problems can be solved with the aid of
a nonlinear integral equation is of independent interest, and will be
put to use in \S\ref{bbsec} below.

\resection{The harmonic oscillator and the free-fermion point}
\label{shosec}
The case $M=1$ falls just outside the `semiclassical domain' $M>1$,
$\beta^2<1/2$ that we have been discussing. However, it corresponds
to the three-dimensional harmonic oscillator, for which the exact
wave-functions are known. As explained in, for example, \cite{GP1},
the general solution to the radial Schr\"odinger
equation for the harmonic oscillator at angular
momentum $l$ can be written in terms of the confluent hypergeometric
functions $M(a,b,z)$ and $U(a,b,z)$ (the notation is as 
in~\cite{AS1}). The correctly-normalised solution, subdominant
in the sector ${\cal S}_0$, turns out to be
\eq
y(x,E,l)=
x^{l+1}e^{-x^2/2}U\!\lf(\fract{1}{2}(l{+}\fract{3}{2}){-}\fract{E}{4},
l{+}\fract{3}{2}, x^2\ri)
\label{monesol}
\en
and has the asymptotic behaviour
\eq
y(x,E,l)\sim x^{-1/2+E/2}[1+O(x^{-2})]\exp(-\fract{1}{2}x^{2})\,.
\label{modasymp}
\en
(Notice the extra factor of $x^{E/2}$ here compared to the earlier
formula (\ref{asrep}), reflecting the fact that the semiclassical domain
has
been left.) Taking from \cite{AS1} the analytic continuation formula
\eq
U(a,b,ze^{2\pi in})=\bigl(1{-}e^{-2\pi i b n}\bigr)
\frac{\Gamma(1{-}b)}{\Gamma(1{+}a{-}b)}M(a,b,z)+
e^{-2\pi i b n} U(a,b,z)
\en
and the Wronskian
\eq
W[U(a,b,z),M(a,b,z)]=\frac{\Gamma(b)}{\Gamma(a)}z^{-b}e^z\,,
\en
the function
$\frac{1}{2i}W_{-1,1}(E,l)$ can be calculated explicitly. The 
result is
\eq
\fract{1}{2i}W_{-1,1}(E,l)\Big|_{M{=}1}=\frac{2\pi}{
\Gamma\lf(\fract{3}{4}{+}\fract{l}{2}{+}\fract{E}{4}\ri)
\Gamma\lf(\fract{1}{4}{-}\fract{l}{2}{+}\fract{E}{4}\ri)}~.
\label{Cformula}
\en
At $M{=}1$ we have $\nu=1/\sqrt{4\pi}$, $E=4\pi\lambda^2$, and $l=4p{-}1/2$, 
and so from (\ref{twrel}) we find
\eq
T(\lambda,p)\Big|_{\beta^2{=}1/2}=\frac{2\pi}{
\Gamma\lf(\fract{1}{2}{+}2p{+}\pi\lambda^2\ri)
\Gamma\lf(\fract{1}{2}{-}2p{+}\pi\lambda^2\ri)}~.
\en
This agrees with the result obtained by the authors of
 \cite{BLZ2} for the vacuum
expectation value of the operator
${\bf T}$ at $\beta^2=1/2$, the so-called free-fermion point, for
the particular value $0$ of their non-universal renormalisation constant
${\cal C}$. 

One subtlety has been glossed over here. The modified 
asymptotic (\ref{modasymp}) changes the value of $W_{0,1}(E,l)$
from $2i$ to $2i\,e^{\pi i E/4}$, so that $\tilde C(E,l)$ is no
longer equal to $-1$, but rather
$-e^{-\pi iE/2}$. Hence (\ref{tqyrel}) becomes
\eq
C(E,l)y_0(x,E,l)=y_{-1}(x,E,l)+e^{-\pi i E/2}y_1(x,E,l)
\en
and the would-be T-Q relation (\ref{cdeq}) must be replaced by
\eq
e^{\pi i E/4}C(E,l)D^{\mp}(E,l)=
\omega^{\mp\pi i(l+1/2)+E/2}D^{\mp}(\omega^{-2\!}E,l)+
\omega^{\pm\pi i(l+1/2)-E/2}D^{\mp}(\omega^{2\!}E,l)
\en
with $\omega=e^{\pi i/2}$. This precisely matches
the `renormalised' T-Q
relation obtained in \cite{BLZ2} for the free-fermion point, so long
as $T(\lambda,p)$ is identified
with $e^{\pi iE/4}\,C(E,l)$ at this point
rather than $C(E,l)$. But when this $T$ is reexpressed in
terms of Wronskians, 
the new prefactor
is exactly cancelled by
the modification
to $W_{0,1}(E,l)$,
with the result that the general relation 
(\ref{twrel}) between $T$ and $W_{-1,1}$ survives unscathed.
Considering the asymptotics discussed in appendix A, we expect
similar modifications to the T-Q relation to occur at $M=1/(2m{-}1)\,$,
$\beta^2=1-1/2m\,$, for any positive integer $m$.

Now we can return to property (iii) from the 
last section, which states that for $l\in (-1{-}M/2,M/2)$, there are no
zeroes of $C$ on the positive-real $E$ axis. 
We will only discuss $M\ge 1$. Then a WKB analysis as in \cite{BB1}
can be used to show that at large $|E|$ all zeroes must lie near the 
negative-real axis, and hence satisfy property (iii).
(In fact, it seems clear that for $M\ge 1$ they lie {\em on}
the negative axis, but this is harder to prove.)
There remains the possibility that some zeroes at smaller values of $|E|$
lie on the positive-real
axis, and to exclude this at least for $M$ near $1$ we use
the fact that the
zero positions at $M{=}1$ are determined by
(\ref{Cformula}), and are 
\eq
E=-4n-2\pm (2l+1)~,\qquad n=0,1,\dots
\label{cvalpos}
\en
Thus for $M{=}1$ and
$-3/2<l<1/2$ the zeroes are bounded away from the positive-real axis.
By uniform 
continuity in compact domains, this will also to be true for the
low-lying zeroes if $M$ is sufficiently close to $1$.
(Note that continuity alone does not exclude the possibility of a zero 
at large $|E|$ violating (iii) for $M$ arbitrarily near to $1$, which
is why we also had to invoke the WKB argument.)
In fact, it appears that all of the zeroes of $C$
actually lie on the negative-real 
axis for $M\ge 1$ and $l\in(-1{-}M/2,M/2)$. 
At $l=0$, this is the main conjecture of \cite{BB1}; some
numerical evidence for the claim at other 
values of $l$ is given in \S\ref{bbsec}. However,
the given range for $l$ is certainly maximal: from
the formula
\eq
C(0,l)=2\cos\frac{2l{+}1}{2M{+}2}\pi\,,
\label{czform}
\en
a consequence of the T-Q relation (\ref{cdeq}), it follows that
$C(E,l)$ has a zero at $E{=}0$ 
when $l=-1{-}M/2$ and $l=M/2$.
 
Finally in this section we calculate $D^-(0,l)\bigl|_M\,$. At
$M{=}1$\,, (\ref{monesol}) can be used with the relation
$U(a,2a,x^2)= \frac{1}{\sqrt{\pi}}
x^{1-2a}
e^{x^2/2} K_{a-1/2}(x^2/2)\,,$
where $K_a$ is a Bessel function of second kind, to see that
\eq
y(x,0,l)\bigl|_{M{=}1}={\,1\over \sqrt{\pi}}\,x^{1/2}K_{l/2+1/4}(x^2/2)\,.
\en
For $\nu>0$ and $z\rightarrow 0$,
$K_{\nu}(z)\sim\frac{1}{2}\Gamma(\nu)\,(z/2)^{-\nu}$ 
\cite{AS1}\,, so for $l>-1/2$ we have
\eq
y(x,0,l)\bigl|_{M=1}\, \sim 
{\,1\over\sqrt{\pi}}\,2^{l-1/2}\, 
\Gamma (\fract{l}{2} {+} \fract{1}{4})\,x^{-l}
\qquad \mbox{as~}x\rightarrow 0\,,
\en
and hence, using (\ref{daltform}), $D^-(0,l)\bigl|_{M=1}\,=
\frac{\,1}{\sqrt{\pi}}2^{l+3/2}\,\Gamma(1{+}\frac{l}{2}{+}\frac{1}{4})\,$.

The general-$M$ case can be recovered by
using the fact that at $E{=}0$ a variable change relates 
a solution at arbitrary $M$ and $l$ to a solution
at $M{=}1$, but with $l$ replaced by
$l'=l'(l,M)$.
Considering that
the normalisation at large $x$ must be in agreement with (\ref{asrep}),
the relevant transformation is  
\eq
y(x,0,l)\bigl|_M\,= 
( \fract{2}{M+1})^{{1}/{4}} x^{-(M{+}1)/4}  
y( (\fract{2}{M+1})^{1/2} x^{(M{+}1)/2}, 0,
\fract{2l+1}{M+1}-\fract{1}{2})\bigl|_{M=1}
\label{transf}
\en
and repeating the steps already described for $M{=}1$, the
result quoted in the last section is recovered:
\eq
D^-(0,l)\bigl|_M\,=
(\alpha^-)^{-1}=
\fract{\,1}{\sqrt{\pi}} 
\Gamma(1+\fract{2l+1}{2M+2})\,
(2M{+}2)_{\phantom{+}}^{\frac{2l+1}{2M+2}+ \frac{1}{2}}\,.
\label{dmres}
\en
To find $D^+(0,l)$, we continue from
$l$ to $-1{-}l\,$:\\[-4pt]
\eq
D^+(0,l)\bigl|_M\,=
(\alpha^+)^{-1}=
\fract{\,1}{\sqrt{\pi}} 
\Gamma(1-\fract{2l+1}{2M+2})\,
(2M{+}2)_{\phantom{+}}^{-\frac{2l+1}{2M+2}+ \frac{1}{2}}\,.
\label{dpres}
\en
Notice that
\eq
D^-(0,l)\bigl|_M
D^+(0,l)\bigl|_M\,=
(\alpha^-\alpha^+)^{-1}=
(2l{+}1)/{\sin\fract{2l+1}{2M+2}\pi}\,.
\label{proddet}
\en
When $l{=}0$, $D^-$ and $D^+$ correspond to
half-line problems with 
Dirichlet and Neumann boundary conditions at the origin
respectively,
and therefore yield
the odd and even spectral determinants 
for the whole-line
problem (\ref{Sch}).
(This is why the $\pm$ signs were 
swapped in our definitions
between $D^{\pm}$ and $A_{\mp}$.)
Their product is the full spectral determinant, and 
the result (\ref{proddet}) 
at $l{=}0$
matches the formula for $D_M(0)$ given in \cite{V5,V6}.

\blank{
then we get
\eq
(2l+1) y(x,l) \sim  {x^{-l} \over   \alpha^-(M, l)} 
\en
with
\eq
\alpha^{-} (M,l)=  { \sqrt{\pi} \over  \Gamma ( 1 + \fract{ 2l+1}{2M+2} )}
(2+2M)^{- \fract{2l+1}{ 2+2M} - \fract{1}{2} } 
\label{alp1}
\en
furthermore
\eq
\alpha^{+} (M,l)= \alpha^{-} (M,-l-1)
\label{alp2}
\en
Notice that 
\eq
\alpha^{+} (M,l)\alpha^{-}(M,l)= \sin (\pi \fract{2l+1}{2 M+2})/(2l+1)
\en
hence~(\ref{DD1}) is transformed into~(\ref{AA}).
Let's stress that~(\ref{transf}) allow to express  the  solution for the
linear potential at $l=0$ in terms of Bessel functions, 
thus we have the well known equivalence
\eq
\pi \Ai(x) =    \sqrt{ x \over 3 } K_{1/3}( \fract{2}{3} x^{3/2})
\en
}

\resection{Fusion relations and monodromy}
\label{monodsec}
The T-Q relation is not the only functional equation which arises in the
context of
the ${\bf T}$- and ${\bf Q}$-operators (see, for 
example,~\cite{KP,KNSa,BLZ1,FLSa,BLZ2,BLZ3}). Further relations are
conveniently expressed using the `fused' ${\bf T}$-operators ${\bf
T}_j$, which can
be built up by a process known as fusion from the basic operator ${\bf
T}$. Introducing a half-integer valued 
index $j=0$, $1/2$, $1,\dots$,
the first set of fusion relations, sometimes called a T-system,
reads as follows:
\eq
{\bf T}_j(q^{-1/2}\lambda)
{\bf T}_j(q^{1/2}\lambda)
={\bf 1}+
{\bf T}_{j+1/2}(\lambda)
{\bf T}_{j-1/2}(\lambda)\,
\label{tsys}
\en
where ${\bf T}_0(\lambda)\equiv {\bf 1}$ and
${\bf T}_{1/2}(\lambda)\equiv {\bf T}(\lambda)$. 
The fused ${\bf T}$'s can also be obtained from
\eq
{\bf T}(\lambda)
{\bf T}_j(q^{j+1/2}\lambda)
=
{\bf T}_{j-1/2}(q^{j+1}\lambda)
+
{\bf T}_{j+1/2}(q^j\lambda)\,
\label{fusone}
\en
or
\eq
{\bf T}(\lambda)
{\bf T}_j(q^{-j-1/2}\lambda)
=
{\bf T}_{j-1/2}(q^{-j-1}\lambda)
+
{\bf T}_{j+1/2}(q^{-j}\lambda)\,.
\label{fustwo}
\en
The vacuum states $|p\rangle$ are also eigenstates of
these fused ${\bf T}$-operators. In this section we shall show that
the vacuum
expectation values $T_{j}(\lambda)\equiv 
\langle p|{\bf T}_j(\lambda)|p\rangle$ arise naturally in the context
of the Schr\"odinger equation~(\ref{sibeq}), leading to 
a reinterpretation of
the fusion relations and their truncations in terms of the behaviour
of solutions to this equation under analytic continuation.
 
As remarked earlier, each pair of functions $\{y_m,y_{m+1}\}$ provides
a basis for the space of solutions to (\ref{sibeq}). So far,
we have only examined the expansion of $y_{k-1}$ in the basis
$\{y_k,y_{k+1}\}$, but it is natural to ask about other
possibilities. To this end, we extend the definition (\ref{stokedef})
of $C_k$ and $\tilde C_k$ by setting
\eq
y_{k-1}=C^{(m)}_ky_{k+m-1}+\tilde C^{(m)}_ky_{k+m}
\label{fusedstokes}
\en
(so that $C_k^{(1)}=C_k$ and $\tilde C_k^{(1)}=\tilde C_k=-1\,$). 
The change from the $\{y_{k+m-1},y_{k+m}\}$ basis
to the $\{y_{k-1},y_{k}\}$ basis is then effected by a $2\times 2$
matrix ${\bf C}^{(m)}_k$ as 
\eq
\lf(\matrix{y_{k-1}\cr y_{k}}\ri)=
{\bf C}^{(m)}_k\lf(\matrix{y_{k+m-1}\cr y_{k+m}}\ri)~~,
\quad
{\bf C}^{(m)}_k=
\lf(\matrix{C^{(m)}_k&\tilde C^{(m)}_k\cr
            C^{(m-1)}_{k+1}&\tilde C^{(m-1)}_{k+1}}\ri)\,.
\en
This matrix depends on $E$ and $l$, but not $x$.
The following properties are immediate:
\eq
{\bf C}^{(m)}_k(E,l)=
{\bf C}^{(m)}_{k-1}(\omega^2E,l)~,
\en

\eq
{\bf C}^{(0)}_k=
\lf(\matrix{1&0\cr
            0&1}\ri)~~,\quad
{\bf C}^{(1)}_k=
\lf(\matrix{C_k&-1\cr
            1&0}\ri)~.
\en
Further relations
reflect the fact that the change from the basis 
$\{y_{k+m+n-1},y_{k+m+n}\}$ to 
$\{y_{k+m-1},y_{k+m}\}$,
followed by the change from
$\{y_{k+m-1},y_{k+m}\}$ to 
$\{y_{k-1},y_{k}\}$, has the same effect as
accomplishing the overall change in one go: 
\eq
{\bf C}^{(m)}_k
{\bf C}^{(n)}_{k+m}
={\bf C}^{(m+n)}_{k}\,.
\label{monrel}
\en
(These express the consistency of the analytic continuations, and can
be thought of as monodromy relations.) Consider first the case
$m=1$. We have
\eq
\biggl(\matrix{C_k&-1\cr
            1&0}\biggr)
\biggl(\matrix{C^{(n)}_{k+1}&\tilde C^{(n)}_{k+1}\cr
            C^{(n-1)}_{k+2}&\tilde C^{(n-1)}_{k+2}}\biggr)=
\biggl(\matrix{C^{(n+1)}_{k}&\tilde C^{(n+1)}_{k}\cr
            C^{(n)}_{k+1}&\tilde C^{(n)}_{k+1}}\biggr)\,,
\en
which gives two non-trivial relations:
\eq
C^{\phntn}_k C^{(n)}_{k+1}-C^{(n-1)}_{k+2}=C^{(n+1)}_k
\label{Crel}
\en
and
\eq
C^{\phntn}_k\tilde C^{(n)}_{k+1}-\tilde C^{(n-1)}_{k+2}=\tilde C^{(n+1)}_k.
\label{Ctrel}
\en
In addition, we have the initial conditions
\bea
C^{(0)}_k=1~~&,&\quad C^{(1)}_k=C_k~\\[4pt]
\tilde C^{(0)}_k=0~~&,&\quad \tilde C^{(1)}_k=-1~
\label{Cinit}
\eea
The $n=1$ case of (\ref{Ctrel}) shows that $\tilde
C^{(2)}_k=-C^{\phntn}_k=-C^{(1)}_k\,$; and then the more general equality
\eq
\tilde C^{(n)}_k=-C^{(n-1)}_k
\label{tceq}
\en
follows on comparing (\ref{Ctrel}) with (\ref{Crel}). If we now
set
\eq
C^{(n)}(E)=C_0^{(n)}(\omega^{-n+1}E)\,,
\en
then (\ref{Crel}) is equivalent to
\eq
C(E)C^{(n)}(\omega^{n+1\!}E)=C^{(n-1)}(\omega^{n+2\!}E)+
C^{(n+1)}(\omega^{n\!}E)\,,
\label{c0sys}
\en
and this matches the fusion relation (\ref{fusone}). Since
$C^{(0)}(E)=1=T_0(E)$ and, from the last section,
$C^{(1)}(E)=C(E)=T_{1/2}(\nu E^{1/2})$, this
establishes the basic equality
\eq
C^{(n)}(E)
=T_{n/2}(\nu E^{1/2})~.
\label{beqq}
\en
It is easy to check that the fusion relation
(\ref{fustwo}) emerges in a similar manner
from (\ref{monrel}) at $n=1$. To recover the T-system
(\ref{tsys}), one more piece of information is needed. Taking
Wronskians in (\ref{fusedstokes}) yields
\eq
C^{(m)}_k=\frac{1}{2i}W_{k-1,k+m}~,~~
\tilde C^{(m)}_k=-\frac{1}{2i}W_{k-1,k+m-1}\,.
\label{cwrk}
\en
An immediate consequence is the recovery of (\ref{tceq}), but we also
obtain
\eq
C^{(m)}_k=-C^{(-m-2)}_{k+m+1}~.
\en
Using this result, the $n=-m$ case of (\ref{monrel}), 
namely
${\bf C}^{(m)}_k {\bf C}^{(-m)}_{k+m}=1$, implies that
\eq
C^{(m-1)}(\omega^{-1\!}E)C^{(m-1)}(\omega E)
-C^{(m)}(E)C^{(m-2)}(E)
=1\,.
\label{csys}
\en
Given the identification (\ref{beq}), this is the T-system
(\ref{tsys}), evaluated on the vacuum state $|p\rangle$.
Finally, the formula (\ref{cwrk}) allows the function $T_{n/2}$ to
be expressed alternatively in terms of a Wronskian:
\eq
T_{n/2}(\nu E^{1/2})
=C^{(n)}(E)
=\frac{1}{2i}W_{-1,n}(\omega^{-n+1\!}E)~.
\label{beq}
\en
This will be relevant in \S\ref{bbsec} below.

The monodromy relations used so far have all been `local', in that
they can be built up from continuations of the functions $y_k$
from one sector $\CS_n$ to its
neighbours $\CS_{n\pm 1}$. But there is also the possibility to
continue all the way round the origin a number of times. If the cover
of the punctured $x$-plane on which the $y_k$ live closes in a suitable
sense, this leads to new relations, and these turn out to correspond to
the truncations of fusion hierarchies in the
integrable quantum field theory.

As an illustration,  we will discuss the simplest class of
examples, which arise when $2M$ is an integer and $l(l{+}1)$ is 
zero. Then all
solutions to (\ref{sibeq}), and in particular the 
$y^{\phntpm}_k(x,E,l)$,
are single valued on the once-punctured $x$-plane, and the sector
$\CS_{n+2M+2}$ coincides with the sector $\CS_n$. Both $y^{\phntpm}_n$ and
$y^{\phntpm}_{n+2M+2}$ are subdominant in this sector, and so they must be
proportional to each other. To find the precise relationship, consider
their behaviour as
$|x|\rightarrow\infty$ in $\CS_n$, setting $x=\rho\,
e^{n\pi i/(M{+}1)}$ and letting $\rho\rightarrow\infty$. From
(\ref{asrep}) and (\ref{ykdef}),
\eq
y^{\phntpm}_n\sim \omega^{n/2}\rho^{-M/2}
\exp(-\fract{1}{M+1}\rho^{M+1})~,~~~
y^{\phntpm}_{n+2M+2}\sim -\omega^{n/2}\rho^{-M/2}
\exp(-\fract{1}{M+1}\rho^{M+1})
\en
and so $y^{\phntpm}_{n+2M+2}(x,E,l)=-y^{\phntpm}_n(x,E,l)$. 
{}From (\ref{beq}) we immediately
deduce
\eq
C^{(2M)}(E)=1 ~~~~,~~~~ 
C^{(2M+1)}(E)=0
\en
and hence the
relation~(\ref{csys}) can be rewritten as
\eq
C^{(m)}( \omega^{-1}E)C^{(m)}( \omega E)= 
1+ \prod_{n=1}^{2M-1} \lf ( C^{(n)}(E) \ri)^{l_{nm}} 
\label{oldpa}
\en
where $l_{nm}$ is the incidence matrix of the $A_{2M-1}$ Dynkin diagram. 
This is the simplest example of truncation, and ultimately~\cite{KP}
leads, via
the Y functions discussed briefly in
\S\ref{yssec} below, to alternative sets of integral equations of
the `TBA' type~\cite{Zam1}.
For integer values of $M$, these equations were applied
in~\cite{DT3} to the computation of
the energy levels of anharmonic oscillators with
analytic potentials $x^{2M}$.

At arbitrary rational values of $M{+}1$, a similar collapse of the
set of solutions $\{y_k\}_{k=-\infty}^{\infty}$ occurs, but this time
on some finite cover of the punctured $x$-plane. Again, a 
truncation of the fusion hierarchy results. 
For nonzero values of $l(l{+}1)$ the
situation is a little more subtle, as there is a multivaluedness in
the solutions induced by the singularity at $x=0$. To handle this
behaviour the alternative basis $\{\psi^-,\psi^+\}$ is more appropriate.
We leave further discussion of this point to future work, but we
expect that all of the more general truncations of the fusion
hierarchy (\ref{csys}) will eventually
find a geometrical interpretation in terms of the behaviours of
solutions to the basic differential equation~(\ref{sibeq}).

\resection{The (fused) quantum Wronskians}
\label{tqgenrel}
The next task is to relate
the fused Stokes multipliers $C^{(m)}$ to the
functions $D^{\mp}$ discussed in \S\ref{tqsec}. 
Recall that the Wronskian is   bilinear 
in the  space of differentiable functions, so that
given  
four   arbitrary functions
$f(x)$, $g(x)$, $h(x)$ and $l(x)$,
and arbitrary constants $\alpha,\beta,\gamma,\delta$,
 we have 
\eq
W[\alpha f+\beta g,\gamma h+ \delta l]=\alpha \gamma W[f,h]
+\alpha \delta  W[f,l] 
+ \beta \gamma  W[g,h] + \beta \delta  W[g,l]~~.
\label{wprop}
\en
For almost all $l$ (exceptions will be discussed below),
the functions
$ \{ \psi^- ,  \psi^+ \}$ introduced in (\ref{psidef}) and
(\ref{psidefmin})
provide an alternative basis for the space of solutions to the
differential equation (\ref{sibeq}).
In particular, 
using the results $W[y,\psi^{\pm}]=D^{\mp}$ and
$W[\psi^-,\psi^+]=2l{+}1$, we have
\eq
(2l{+}1)y(x,E,l)=  
D^-(E,l)\psi^-(x,E,l)-D^+(E,l)\psi^+(x,E,l)\,,
\qquad~~ 
\label{yd}
\en
More generally, the shifted solutions defined by (\ref{ykdef}) and
(\ref{shifted}) satisfy
\eq
{~}\quad
(2l{+}1)y^{\phntpm}_k(x,E,l)=  
D^-(\omega^{2k\!}E,l)\psi_k^-(x,E,l)-
D^+(\omega^{2k\!}E,l)\psi_k^+(x,E,l)\,. 
\label{yds}
\en
Taking the Wronskian (\ref{yds}) at $k=-1$ with the same equation at
$k=n$, shifting $E$ to $\omega^{1-n}E$ and then using
the formula (\ref{beq})
for $C^{(n)}(E)$, property~(\ref{wprop}) and the results
\bea
W[\psi_k^+,\psi_p^+] &=& W[\psi_k^-,\psi_p^-]=0~, \nn \\
W[\psi_p^-,\psi_k^+] &=& (2l{+}1) \omega^{(k-p)(l {+} 1/2)}
\eea
(valid at arbitrary `shifts' $p$ and $k$),
we find
\bea
 (4l{+}2)i\, C^{(n)}(E) &=&  
\omega^{(n+1)(l+1/2)} 
D^-(\omega^{n+1\!} E,l)D^+(\omega^{-n-1}E,l) 
\nn\\
&&{~~~~} - 
\omega^{-(n+1)(l+1/2)} 
D^-(\omega^{-n-1} E,l)D^+(\omega^{n+1\!}E,l)~.
\qquad
\label{DD1}
\eea
In the context of integrable quantum field theory, a corresponding
set of relations was given in~\cite{BLZ2}:
\bea
2 i \sin( 2 \pi p) T_j(\lambda)&=&
q^{(4j+2) p/\beta^2} A_+( q^{j+ 1/2}\lambda,p)
A_-(q^{-j-1/2}\lambda,p)\nn\\
&&{~~~~}- q^{-(4j+2) p/\beta^2} A_+( q^{-j-1/2}\lambda,p)
A_-(q^{j+1/2}\lambda,p)  \qquad
\label{AA}
\eea
where  $j=0$, $1/2$, $1\dots\,$.
With the identifications (\ref{adrel}) and (\ref{twrel}), and
using the result (\ref{proddet}),
the two sets of relations agree if, as before,
$q=\omega$
and  $2p/\beta^2 = l+1/2$.
At $j=0$, (\ref{AA}) is the `quantum Wronskian' relation of
\cite{BLZ2}, while the $n=0$ case of
(\ref{DD1}) was first found (for $M$ an
integer and $l=0$) in \cite{V0}. The match between these two 
was a key ingredient in \cite{DT3,BLZschr}.

The `T-system conjecture' of \cite{DT3}, 
proved in \cite{Suz1}, is a simple corollary of this result.
If $l=0$ and  $M$ is an 
integer, then (\ref{DD1}) at $n=M$ becomes
\eq
C^{(M)}(E) = D^+(-E)D^-(-E) = D_M(-E)\,, 
\label{conj}
\en
the second equality following because at $l=0$, $D^+$ and $D^-$
are the even and odd spectral subdeterminants respectively for the
full-line problem (\ref{Sch}).
Since $C^{(M)}(E)= T_{M/2}(\nu E^{1/2})$, 
this establishes the conjectured relation between $D_M$ 
and the vacuum expectation value $T_{M/2}$.  
(Note though the small differences
in notation from \cite{DT3}: as well as the negation of $E$ already
mentioned in the introduction, 
a half-integer $j$ has been used to index
the T operators in this paper,
in line with the conventions 
of \cite{BLZ1,BLZ2,BLZ3},  while in \cite{DT3} the index was
integer-valued. Thus in \cite{DT3}
the correspondence was with $T_M$ rather than
$T_{M/2}$.)

Next we return to the fact that the pair of functions
$\{\psi^-,\psi^+\}$ does not always furnish
a basis for the space of solutions to
the differential equation (\ref{sibeq}). The most obvious counterexample
to such a claim is the point $l=-1/2$, at which $\psi^-=\psi^+$. 
More generally, since $\psi^-$ is initially defined by analytic
continuation in $l$, there may be points at which poles are encountered. If
these poles are removed by multiplying by a regularising factor, the
resulting solution $\bar\psi^-$ may fail to be independent of $\psi^+$
at those values of $l$ where previously there had been divergences.
This is related to the remark in \cite{BLZ2} that at certain
values of $p$ the functions $A_+(\lambda,p)$ and $A_-(\lambda,p)$ may
coincide.
In the context of the radial Schr\"odinger equation, a discussion of
the issue can be found in, for example, chapter~4 of
\cite{NEWT}.
The most direct way to locate the `problem' values of $l$
is probably the iterative
construction of $\psi^-$ and $\psi^+$ given in \cite{HC}. For the
potential $x^{2M}$, for which the function $U(x)$ of \cite{HC} is equal
to $x^{2M}-E$, poles $\psi^+(x,E,l)$ occur at
$l+1/2=-m_1-(M{+}1)m_2$, with $m_1$ and $m_2$ non-negative integers. 
Hence
$\{\psi^-,\psi^+\}$ fails as a basis at the points
\eq
l+1/2=\pm(m_1+(M{+}1)m_2)~, \qquad m_1, m_2\ge 0~.
\label{singpts}
\en
For integer 
values of $2M$, this is just a standard phenomenon in the
Frobenius method, which predicts that one of the pair
$\{\psi^-,\psi^+\}$ may have 
a logarithmic component whenever the two solutions to the indicial equation 
$\alpha(\alpha{-}1)=l(l{+}1)$
differ by an integer, or an even integer when $2M$ is even. 

Rather than giving a complete treatment, we will focus here on 
the $l=-1/2$ case. This corresponds to the 
ground state of the untwisted sine-Gordon model, 
and so is particularly interesting from the field-theoretic
point of view.
The emergence of logarithmic corrections can be understood by taking
an appropriate limit $l \rightarrow  -1/2$.   
Using  (\ref{yd}), valid away from the points (\ref{singpts}),
set $l=-1/2+\varepsilon$ with $\varepsilon$
tending to zero. Near $x=0$ we have
\eq
y(x,E,-\fract{1}{2}+\varepsilon ) \sim 
D^-(E,-\fract{1}{2})'\,x^{1/2}\,
\frac{x^{\varepsilon}{+}x^{-\varepsilon}\!}{2}
\,{}-{}\,
D^-(E,-\fract{1}{2})\, x^{1/2}\,
\frac{x^{ \varepsilon}{-}x^{-\varepsilon}\!}{ 2 \varepsilon}
\label{trouser}
\en   
where the primes denote differentiation with respect to $l$ 
and we 
used the facts that  
$D^-(E,l)$  and $D^-(E,l)'$
are nonsingular at $l=-1/2$ to 
expand $D^{\mp}(E,-1/2{+}\varepsilon) 
\sim D^-(E,-1/2) \pm D^-(E,-1/2)' \varepsilon$, 
Expanding  $x^{\pm \varepsilon}=
\exp( \pm \varepsilon \log x )$ to
first order in  $\varepsilon$, (\ref{trouser}) becomes
\eq 
 y(x,E,-\fract{1}{2}) \sim   D^-(E,-\fract{1}{2})'\, x^{1/2} 
- D^-(E,-\fract{1}{2})\, x^{1/2} \log x \,.
\label{newas}
\en
We see that
$y(x,E,-1/2)$ has an $x^{1/2} \log x$ component at small $x$, as
expected from the Frobenius method.
The basis $\{\psi^-,\psi^+\}$ should therefore be replaced at $l=-1/2$, 
and a suitable choice is
the pair $\{ \chi^-, \chi^+ \}$:
\bea
&&\chi^+(x,E) \sim x^{1/2}  + O(x^{5/2})~, \nn \\
&&\chi^-(x,E) \sim x^{1/2} \log x +O(x^{5/2})~.
\label{newbasis}
\eea
The `Jost functions' (\ref{ddef}) become
\eq
\tilde D^{\mp}(E) = 
W[y(x,E,-\fract{1}{2}),\chi^{\pm}(x,E)]~,
\label{ddefchi}
\en
so that 
$\tilde D^{-}(E)= D^{-}(E,-\fract{1}{2})\,$,
and
$\tilde D^{+}(E)= D^{-}(E,-\fract{1}{2})'\,$. 
It is interesting to see the effect that these changes have on the basic
functional relations. We set
$\chi_k^{\pm}(x,E)= \omega^{k/2} \chi^{\pm}(\omega^{-k} x, \omega^{2k} E)$  
and use the Wronskians 
\bea
W[\chi_p^+,\chi_q^+] &=&  0~, \nn\\
W[\chi_p^-,\chi_q^-] &=&  (q{-}p) \log \omega~,\nn\\
W[\chi_p^+ ,\chi_q^-] &=& 1~,  
\eea
to find, with $\tilde D_k^{\mp} \equiv 
\tilde D^{\mp}( \omega^{2k} E)$,
\eq
W[y_p,y_q]=  (q-p) \tilde D^-_p \tilde D^-_q \log \omega +
   (\tilde D^-_p \tilde D^+_q - \tilde D^-_q \tilde D^+_p )
\en
and feeding this into the formula (\ref{beq}) for $C^{(n)}$
yields
\bea
&&
\!\!\!\!\!\!
C^{(n)}(E)= { \pi (n{+}1) \over 2M{+}2  }  
  \tilde D^-(\omega^{n+1\!}E)\tilde D^-(\omega^{-n-1\!}E)
\label{nDD1}\\
&&\qquad{~~~~~~}+{1\over2i}%
\lf(\tilde D^-(\omega^{-n-1\!}E)\tilde D^+(\omega^{n+1\!}E)
- \tilde D^-(\omega^{n+1\!}E) \tilde D^+(\omega^{-n-1\!} E)
 \ri).\nn
\eea
Note also, from (\ref{dmres}) and (\ref{dpres}), that
\eq
\tilde D^-(0)= \sqrt{ {2M+2 \over \pi}}~,~~~\tilde D^+(0)=
{ 2 \over \sqrt{\pi}} (2{+}2M)^{-1/2} ( \log(2{+}2M) - \gamma_E)
\en
where $\gamma_E \equiv -\Gamma'(1)=0.57721...$ is the Euler-Mascheroni 
constant. 
Finally from (\ref{tqyrel}), (\ref{newas}) and (\ref{newbasis}) 
we obtain
\bea
&&C(E)\tilde D^-(E)  = \tilde D^-(\omega^{-2} E) + 
\tilde D^-(\omega^{2} E) \nn \\[2pt]
&&C(E) \tilde D^+(E) = 
\tilde D^+(\omega^{-2}E)+\tilde D^+(\omega^{2}E)
+
\lf(\tilde D^-(\omega^{-2} E) - \tilde D^-(\omega^{2} E) \ri) 
\log \omega\,.  ~~~~~~~~~
\eea
Thus at $l=-1/2$
one of the two T-Q relations has to be changed. Similarly,
comparing (\ref{nDD1}) with the fused quantum Wronskian 
(\ref{DD1})
shows  that 
the relationship between 
$C^{(n)}(E)$ and $D^{\mp}(E,l)$ undergoes a nontrivial modification.
The  fact that for the ground state energy of the 
untwisted sine-Gordon
model things need to be  slightly  modified was already observed,
from a completely different angle,  in~\cite{CGT}.  There, the
general solution of the sine-Gordon  $Y$-system was described in terms
of a single quasi-periodic function $h(E)$, satisfying
$h(E \omega^{2 M+2}  ) = p+q h(E)$, where $p$ and $q$ can, 
for the purposes of~\cite{CGT} and 
the example investigated here, be taken to be constants.
In order to have the correct UV-limit for
the untwisted sine-Gordon model (corresponding to $l=-1/2\,$),  
a non-vanishing $p$  was required, leading to a $\log(E)$ component in
$h(E)$ (see (\ref{lsi}) below).  In a slightly different 
language, this turns out to be what  we have  derived here. 
A match with the results of~\cite{CGT} 
is the topic  of the next section, where, as 
simple  byproduct of the  mapping,
a curious property of the spectral determinants
is pointed out.

%%%%%%%%%%%%%%%%%%%%%%%%%%%%%%%%%%%%%%%
%
\resection{Y-systems and  dilogarithm identities}
\label{yssec}
As mentioned in \S\ref{monodsec}, there is a second set of functional 
relations, the so-called Y-system,  closely related to the T-system 
discussed in the  previous sections. 
The relation between these two systems is 
\eq
Y_n(E)=C^{n+1}(E) C^{n-1}(E)
\label{Ydef}
\en
and the   $Y$'s fulfil the relation
\eq
Y_n(\omega E)Y_n(\omega^{-1} E)=(1+Y_{n+1}(E))(1+Y_{n-1}(E))
\label{ysystem}
\en
For $M$ integer or half-integer and  $l=0$, 
this system  truncates ($Y_0(E)=Y_{2M}(E)=0$), and 
it coincides  with the  $A_{2M-1}$-related Y-system discussed 
in~\cite{Zam2}. 

On the other hand, the $Y$-functions are related to
the solutions  of TBA equations.
In this framework they 
encode
finite-size effects in integrable quantum field theories, and,
through the consideration of ultraviolet limits, lead
to certain remarkable sum rules for the Rogers dilogarithm function involving
the stationary ($E=0$) solutions of the system~(\ref{ysystem}). 
For $M$ integer and $l=0$, for example,  
the relevant sum rule is
\eq
{ 6 \over \pi^2} \sum_{n=1}^{2M-1} 
L \lf({1 \over1+Y_n(0) }\ri) 
= {2M{-}1 \over M{+}1}  = 
c_{UV}
\label{cuv}
\en
where $c_{UV}$ is the central charge of the $Z_{2M}$ parafermionic 
conformal field theory,
$L(x)$ is  the Rogers dilogarithm 
\eq
L(x)= - {1 \over 2} \int_0^x dy \lf[ {\log(y) \over 1{-}y} + 
{\log(1{-}y) \over y} \ri]~,
\en
and  the values of the constants $Y_n(0)$ involved 
in~(\ref{cuv}) are
\eq
Y_n(0)= {\sin( \pi \fract{n+2}{2M+2}) \sin (\pi  \fract{n}{2M+2}) 
\over   \sin^2( \fract{\pi}{2M+2})  }~~. 
\en 
With some additional complications, sum rules similar to~(\ref{cuv})
can be written for any rational $M$~\cite{kirillov} and  
arbitrary~$l$.   
In~\cite{GT1}  a generalisation of~(\ref{cuv}) 
involving the $E$-dependent $Y$-functions 
was proposed. For an
arbitrary solution $Y_n(E)$ to (\ref{ysystem}) (but again
with $M$ an integer and $l=0\,$) the result is
\eq
{ 6 \over \pi^2} \sum_{n=1}^{2M-1} \sum_{k=0}^{2M+1} L\lf({ 1 
\over 1+Y_n(\omega^k E ) }\ri)  =  2(2M{-}1)\,.
\label{sum}
\en
Dilogarithms also
appear in certain volume calculations in 
three-dimensional  manifolds~(see for example~\cite{Zagier1}), and
related to this idea 
is the fact~\cite{GT2} that
the  general
solution to~(\ref{ysystem}) 
can be expressed using  cross-ratios
\eq
(a,b,c,d)= { (a-c)(b-d) \over (a-d) (b-c) }~.
\en
Surprisingly, the points involved in the cross-ratio can be 
expressed in terms of a single quasi-periodic function of $E\,$:
if $h(E)$ satisfies
\eq 
h(E \omega^{2 M+2}  ) \equiv p+q h(E)
\label{qper}
\en
then
\eq
Y_n(E)=-(h( \omega^n E) 
, h(\omega^{-n-2}E) , 
h( \omega^{-n}E) , h(\omega^{n+2}E) )
\label{crdef}
\en
solves (\ref{ysystem}).

Now we would like to show that the function $h(E)$ is a natural object
in the context of the Schr\"odinger equation. Since $C^{(0)}(E)=1$
for all $E$, we can rewrite $C^{(n{+}1)}(E)$ as
$C^{(n{+}1)}(E)/C^{(0)}(\omega^{n{+}1\!}E)$ and, for generic values of
$l$,  expand out numerator and
denominator using (\ref{DD1}) to find
\eq
C^{(n+1)}(E) =  {\omega}^{(l+1/2)(n+1)} 
{D^+(\omega^{-n-2}E,l) 
\over
D^+(\omega^{n}E,l) }
{
\lf ( k( \omega^{n+2} E) - k( \omega^{-n-2} E) \ri)
\over 
\lf ( k( \omega^{n+2} E) - k( \omega^{n} E) \ri)
}
\en
where
\eq
k(E)= E^{l+1/2} D^{-}(E,l)/D^{+}(E,l)~.
\label{lnons}
\en
Similarly, 
\eq
C^{(n-1)}(E)={\omega}^{-(l+1/2)(n+1)} {  
D^+(\omega^{n}E,l)   
\over
D^+(\omega^{-n-2}E,l)  }
{ \lf ( k( \omega^{n} E) - k( \omega^{-n} E) \ri)
\over
\lf ( k( \omega^{-n} E) - k( \omega^{-n-2} E) \ri) }
\en
and so, from (\ref{Ydef}),
\eq
Y_n(E)=- {
 k( \omega^{n+2} E) - k( \omega^{-n-2} E) 
\over 
k( \omega^{n+2} E) - k( \omega^{n} E) }
~
{ k( \omega^{-n} E) - k( \omega^{n} E) 
\over
 k( \omega^{-n} E) - k( \omega^{-n-2} E) }~.
\label{Yres}
\en
This was
for generic values of $l$:
the story changes at
those points where $\{\psi^-,\psi^+\}$
fails to be a basis. The first
example is $l=-1/2$, where (\ref{DD1}) is replaced by (\ref{nDD1}).
It turns out that the final result (\ref{Yres}) continues
to hold, so long as the definition (\ref{lnons}) is swapped for
\eq
k(E)=\log E +2 \tilde D^+(E)/\tilde D^-(E)\,.
\label{lsi} 
\en
Comparing (\ref{Yres}) with (\ref{crdef}), the
function $k(E)$ as defined in~(\ref{lnons}) gives a realisation of 
$h(E)$ 
for $p=0$ and $q= e^{2\pi i(l{+}1/2)}$, 
while the definition
(\ref{lsi}) coincides with  the case $q=1$,  $p=2\pi i$. 
Note also that when $l=0$, $k(E)$ is
essentially the (reciprocal of) the alternating, or skew, spectral
determinant discussed in \cite{V5,V6}.
It is 
interesting  that the result~(\ref{sum}) (and its generalisations) 
appears to furnish a 
novel form of sum rule for the spectral determinants of 
Schr\"odinger equations.

\resection{The T operators, spectral determinants, and a generalisation of
a problem considered by Bender and Boettcher}
\label{bbsec}
The result
\eq
T_{n/2}(\nu E^{1/2})=C^{(n)}(E)=\frac{1}{2i}W_{-1,n}(\omega^{-n+1\!}E)
\label{beqw}
\en
from \S\ref{monodsec}
implies that the vacuum expectation values of the ${\bf T}$-operators
also have an interpretation as spectral determinants. The right hand 
side of (\ref{beqw}) vanishes if and only if 
$E$ is such that the functions
$y_{-1}$ and $y_n$ are linearly dependent, which in turn is true if
and only if 
there is a nontrivial
solution to (\ref{sibeq}) which decays to
zero as $x$ tends to infinity
in the sectors $\CS_{-1}$ and $\CS_n$. This is an eigenvalue problem,
and the argument just given shows that the zeroes of $C^{(n)}$
coincide with the zeroes of its spectral determinant.
{}From (\ref{DD1}) and (\ref{Dlim}),
the order of $C^{(n)}$ is less than $1$ for $M>1$, so the  discussion
around (\ref{factor}) for $D^-$ applies equally 
to $C^{(n)}$, and shows that this function is actually equal to the
spectral determinant, with a normalisation which again turns out to
coincide with the natural one for $2M$ an integer and $l$ equal to $0$ or
$1$. (Incidentally, for $M$ integer and $l$ zero this gives us another
easy proof of the T-system conjecture, obtained more indirectly via
the subdeterminants $D^{\pm}$ in \S\ref{tqgenrel} above.)
General eigenproblems of this sort, where the boundary conditions are
specified 
as $|x|\rightarrow\infty$
in two sectors in the complex $x$ plane,
 are discussed in chapter~6
of \cite{Si}  and also in~\cite{BB1,BBN}. They are 
the other natural set of spectral problems 
associated with (\ref{sibeq}), and it is pleasing
that they correspond so neatly with the fused T-operators. 

As an application, we consider and generalise
a problem
treated by Bender and Boettcher in~\cite{BB1}. These authors
discussed the spectrum of the Hamiltonian
$p^2-(ix)^N$ with $N$ real, corresponding to the eigenvalue equation
\eq
-\psi''(x)-(ix)^N\psi(x)=E\psi(x)\,.
\label{bbprob}
\en
For quantised energy levels the boundary conditions must be chosen
appropriately. The standard requirement is for $\psi(x)$ to tend to 
zero as
$|x|\rightarrow\infty$ on the real axis, but for arbitrary $N$ 
the problem needs to be continued into the complex $x$-plane. The real
$x$-axis is replaced by some contour, and the boundary conditions are
imposed at the two ends of this contour. 
The regions where $\psi(x)$
can vanish exponentially as $|x|\rightarrow\infty$ are wedges, bounded
by what Bender and Boettcher referred to as Stokes lines, with the
vanishing being most rapid at the centers of the wedges, called 
by them anti-Stokes lines. (Beware that many authors use exactly the
opposite terminology -- see, for example, \cite{berry}.)
There are thus a variety of eigenvalue problems associated with
(\ref{bbprob}), depending on the choice of asymptotic directions for
the contour relative to the wedges. The problem studied in \cite{BB1}
was the analytic continuation of the usual harmonic oscillator, so the
wedges were picked to allow the quantisation contour to run along the
real axis when $N=2$.
This selects the
wedges centred on the directions
\eq
\theta_{\rm left}=-\pi+\frac{N{-}2}{N{+}2}\frac{\pi}{2}\quad{\rm and}
\quad
\theta_{\rm right}=-\frac{N{-}2}{N{+}2}\frac{\pi}{2}~,
\label{wedgedef}
\en
each of opening angle $\Delta=2\pi/(N{+}2)$. Bender and Boettcher studied
the problem
both numerically and analytically, and found strong
evidence for an entirely real spectrum for $N\geq 2$, with a transition at
$N=2$ below which infinitely-many eigenvalues become complex. This they
associated with a spontaneous breaking of ${\cal PT}$ symmetry in the
quantum-mechanical problem.

If we set $\hat\psi(x)=\psi(x/i)$, (\ref{bbprob}) becomes
\eq
-\hat\psi''(x)+x^N\hat\psi(x)=-E\hat\psi(x)\,.
\label{nbbprob}
\en
and matches our original Schr\"odinger equation (\ref{sibeq}) 
with $M=N/2$, $l=0$, and
$E$ replaced by $-E$. Furthermore, it is easily checked that the
wedges defined by (\ref{wedgedef}) become the sectors $\CS_{-1}$ and
$\CS_1$. 
Referring back to the discussion following
(\ref{beqw}), this means that the eigenvalues of
the problem posed by (\ref{bbprob}) and (\ref{wedgedef}) 
occur at the negated zeroes of $C^{(1)}(E)$, and are thus encoded
in the pattern of zeroes of the vacuum expectation value of the
fundamental quantum transfer matrix ${\bf T}(\lambda)\equiv
{\bf T}_{1/2}(\lambda)$. 

\begin{figure}[ht]
\vspace{-1.1cm}
\centerline{\epsfxsize=.74\linewidth\epsfysize=.74\linewidth%
\epsfbox{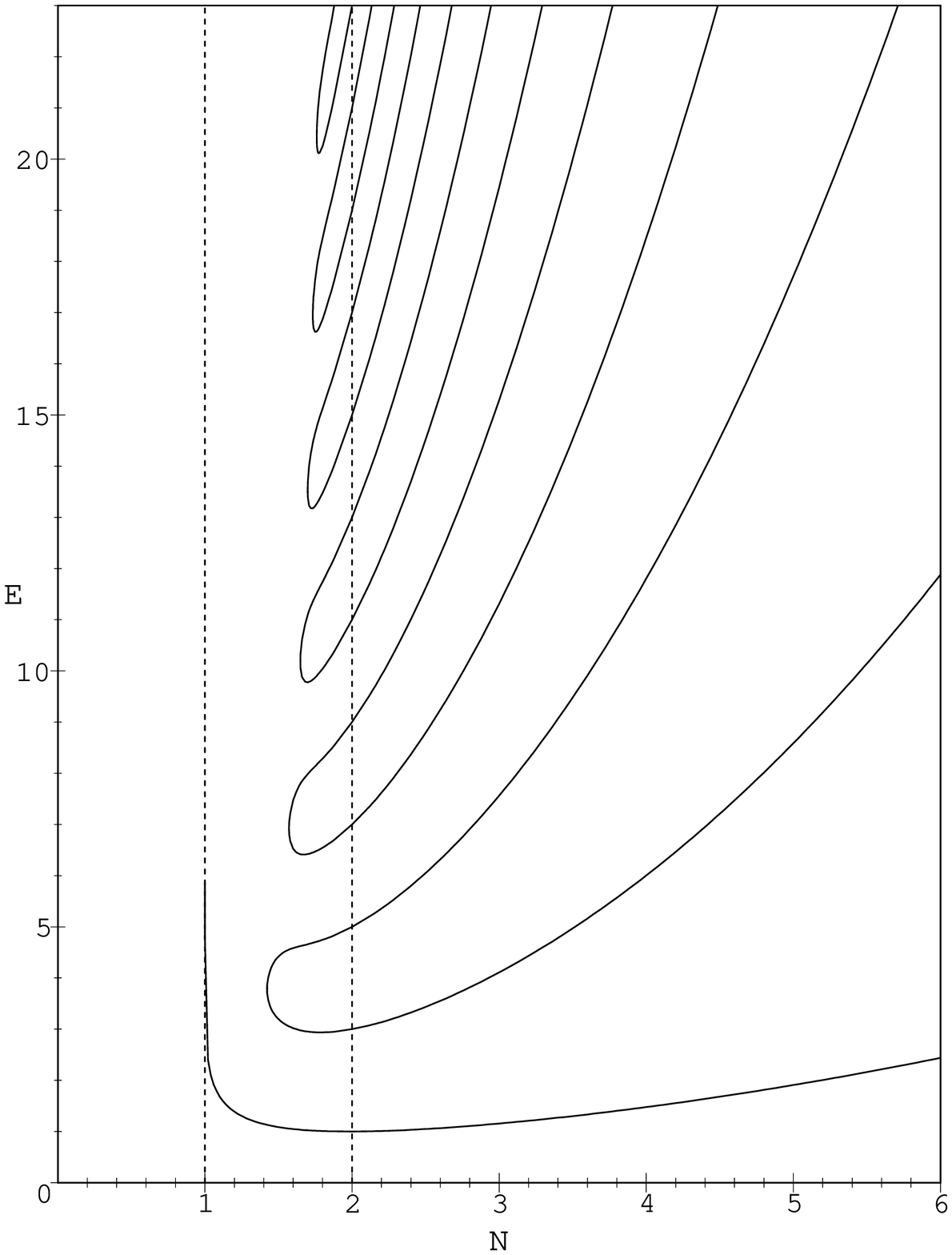}}
\vspace{-1.2cm}
\centerline{\tiny 1) $l=0$\hspace{6.5cm}}
\vspace{.4cm}
\noindent {\small Figure 1: Eigenvalues of the Hamiltonian $p^2-(ix)^N$ plotted
as a function of $N$, found 
via the solution of the nonlinear integral equation (\ref{nlie}) at $l=0$.}
\label{fig1}
\end{figure}

In turn, these can be found via the non-linear integral 
equation (\ref{nlie}).
The results are displayed in figure~1, and agree perfectly with
the results of \cite{BB1}, which were obtained by a direct treatment of the
differential equation in the complex plane.
Note that the transition at
$N=2$ corresponds to the point at which the 
associated sine-Gordon model moves from the attractive to
the repulsive regime, and that the match with the results of Bender
and Boettcher continues to hold even after this point has been passed,
and the semiclassical domain $N>2$ has been left. 

{\begin{figure}
\vspace{-1.5cm}
\[\begin{array}{ll}
\epsfxsize=.45\linewidth\epsfysize=.45\linewidth\epsfbox{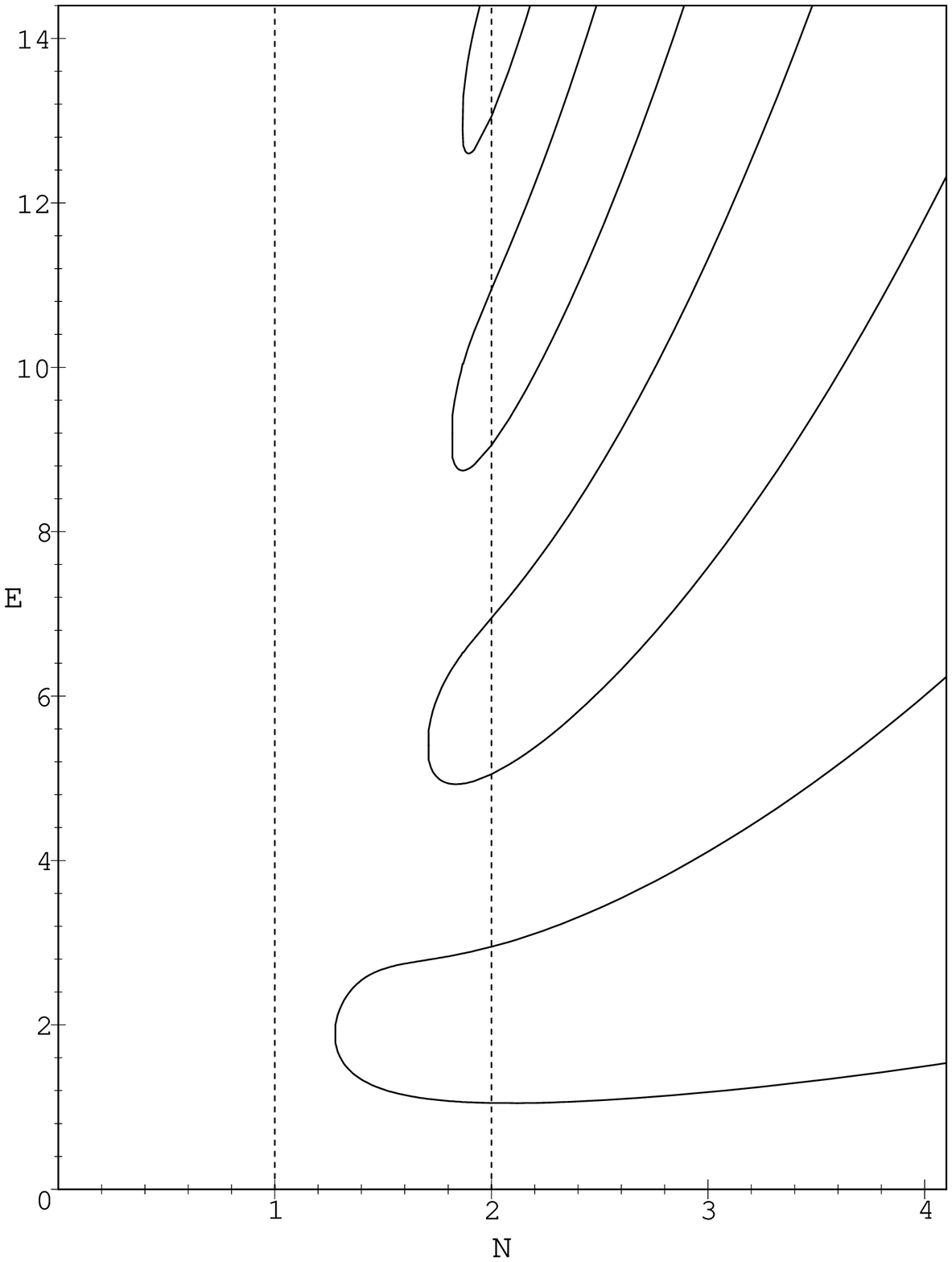}
{}&
\epsfxsize=.45\linewidth\epsfysize=.45\linewidth\epsfbox{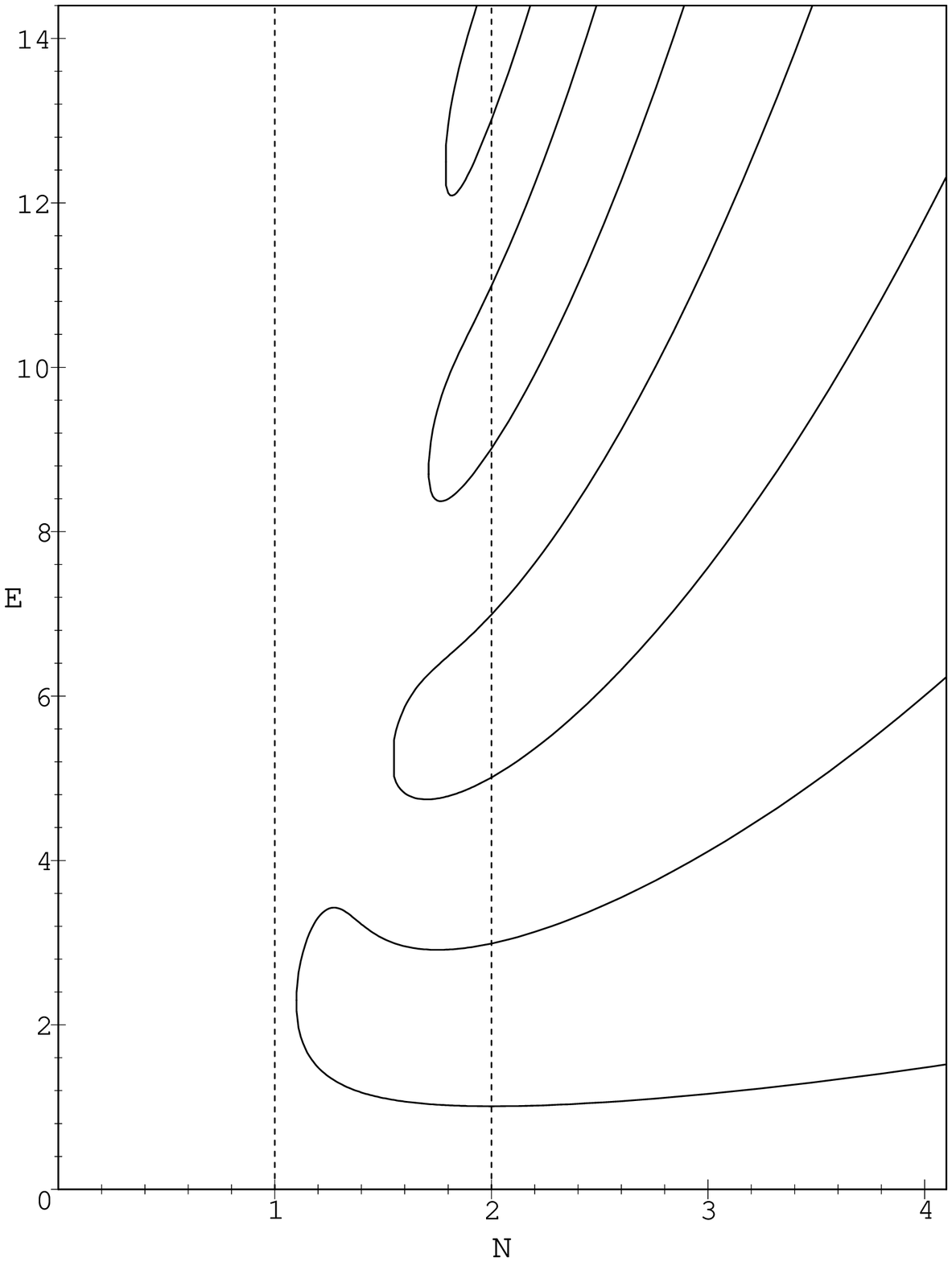}
\\[-19pt]
\parbox[t]{.4\linewidth}{\quad~~~\tiny 2a) $l=-0.025$}
{}~&~
\parbox[t]{.4\linewidth}{\quad~~\tiny 2b) $l=-0.005$}
\\[-7pt]
\epsfxsize=.45\linewidth\epsfysize=.45\linewidth\epsfbox{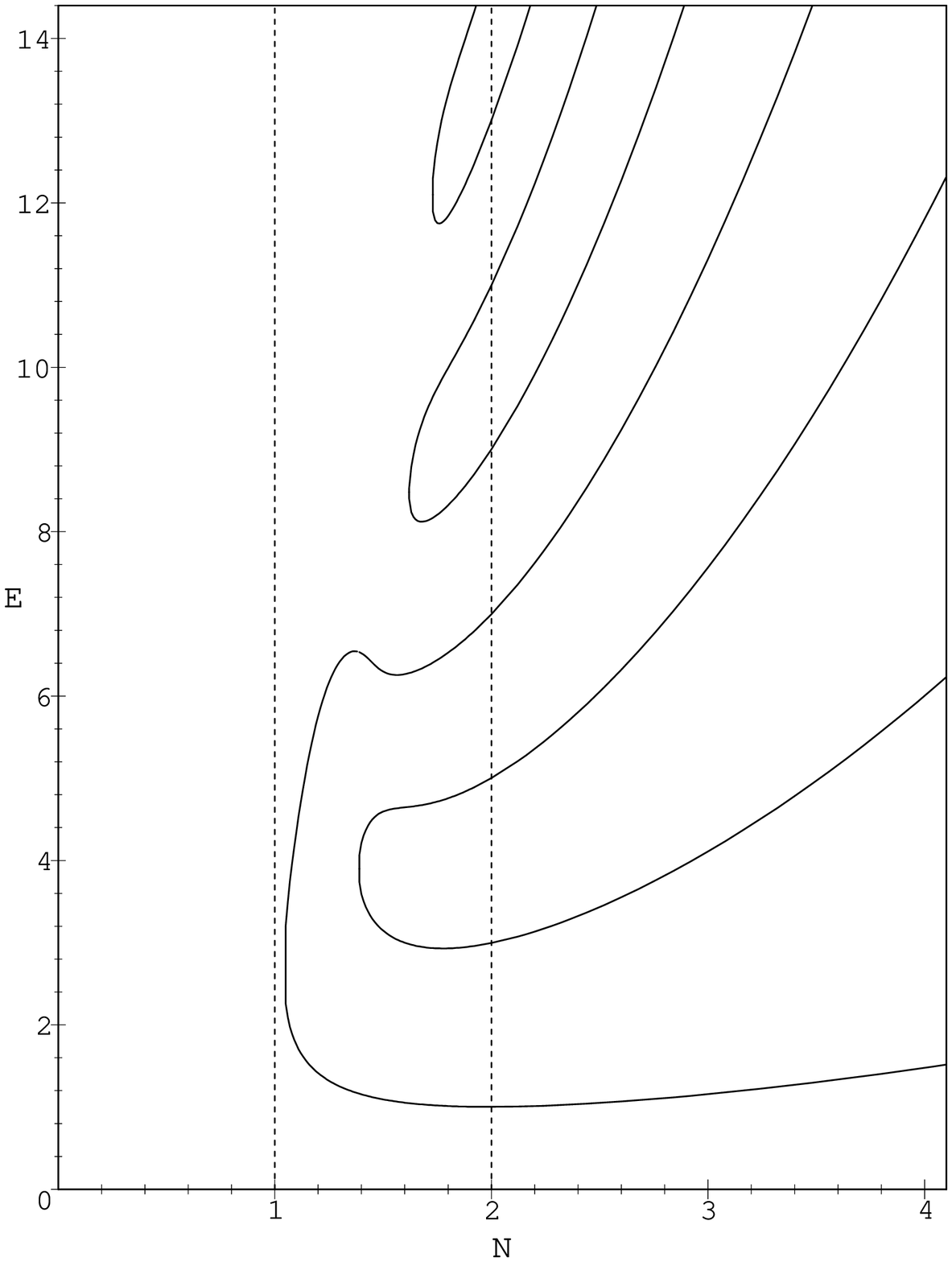}
{}&
\epsfxsize=.45\linewidth\epsfysize=.45\linewidth\epsfbox{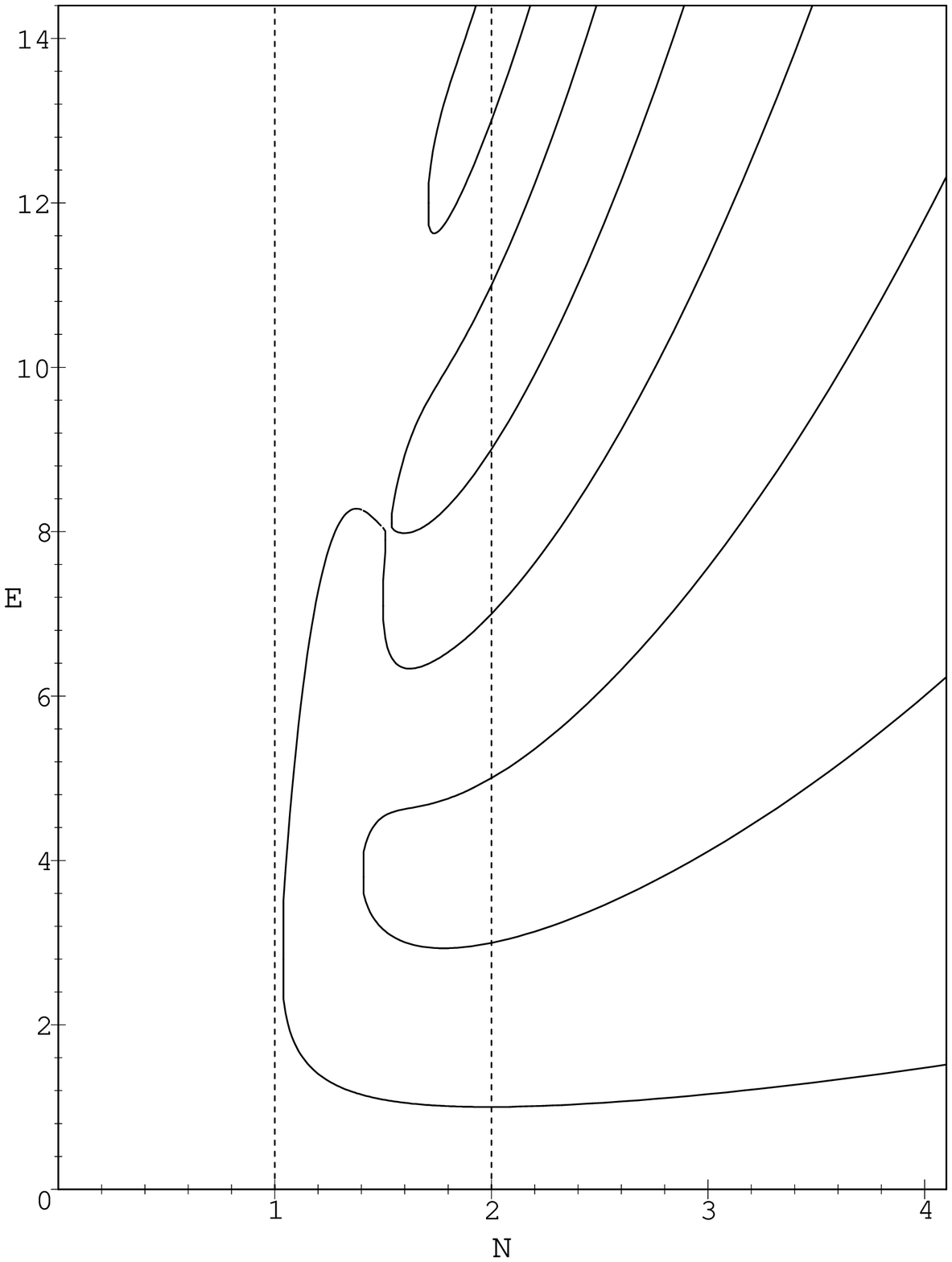}
\\[-19pt]
\parbox[t]{.4\linewidth}{\quad~~~\tiny 2c) $l=-0.0015$}
{}~&~
\parbox[t]{.4\linewidth}{\quad~~\tiny 2d) $l=-0.001$}
\\[-7pt]
\epsfxsize=.45\linewidth\epsfysize=.45\linewidth\epsfbox{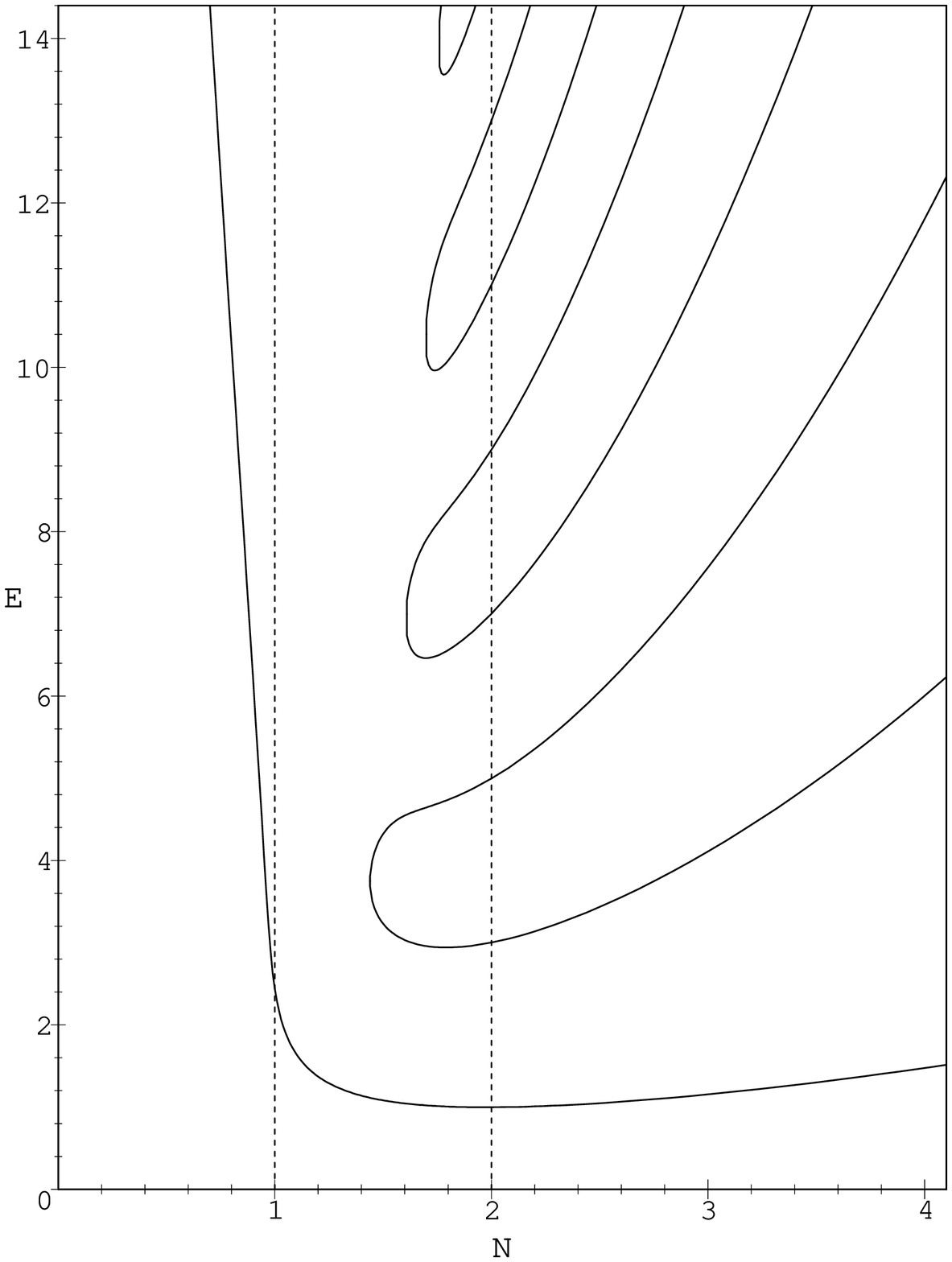}
{}&
\epsfxsize=.45\linewidth\epsfysize=.45\linewidth\epsfbox{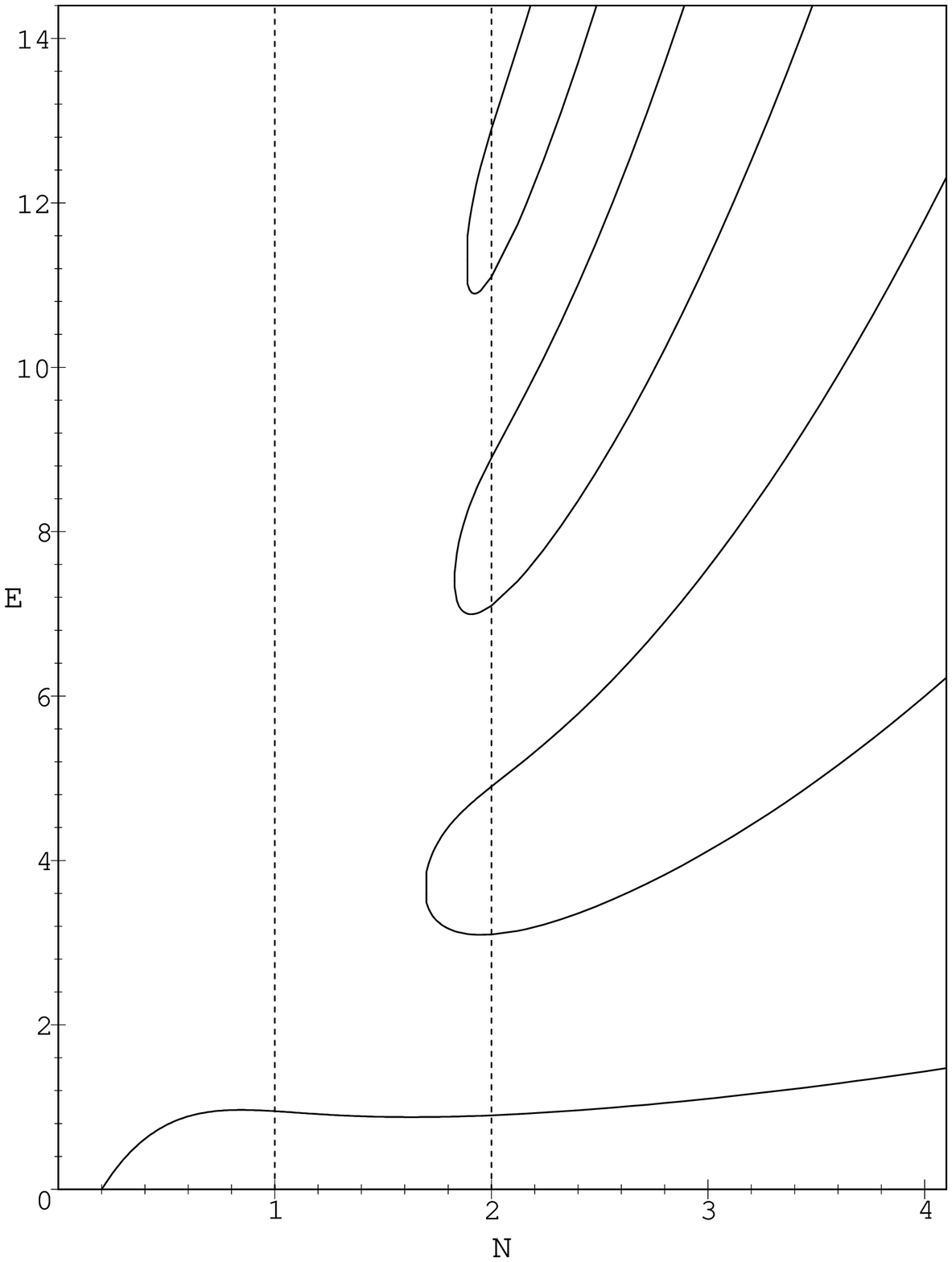}
\\[-19pt]
\parbox[t]{.4\linewidth}{\quad~~~\tiny 2e) $l=0.001$}
{}~&~
\parbox[t]{.4\linewidth}{\quad~~\tiny 2f) $l=0.05$}
\\
\end{array}\]
\noindent {\small Figure 2: Eigenvalues of the Hamiltonian
$p^2-(ix)^N+l(l{+}1)x^{-2}$
found via the nonlinear integral equation (\ref{nlie}),
plotted as a function of $N$ at various nonzero values of $l$.}
\end{figure}}

{}From the point of view of equation (\ref{nlie}), the restriction to
$l=0$ is unnecessary. Once it is relaxed, we are led to 
the following
generalisation of the Bender-Boettcher problem:
\eq
-\psi''(x)-(ix)^N\psi(x)+l(l{+}1)x^{-2}=E\psi(x)\,.
\label{gbbprob}
\en
This seemingly-innocent modification to the Hamiltonian turns out to
have a remarkable effect on its spectrum. 
In figure~2 we show some
results from an initial numerical investigation.
A few features merit immediate comment. 
At $N=2$, the exact eigenvalues can always be found by
negating formula (\ref{cvalpos}). The
behaviour for $N<2$ depends significantly on the sign of $l$.
The situation at $l=-0.025$ is shown in figure 2a.
It might appear similar to that at $l=0$,
but the connectivity of the
states on the plots is exactly reversed: 
the ground state is now joined to the first excited state, 
the second to the third, and so on. As $l$ increases,
the level joined to the ground state flicks up through the rest of the
spectrum, leaving a restored connectivity pattern in its wake. 
The mechanism should be clear enough from figures 2b to 2d.
By the time $l$ reaches zero the earlier situation has been
recovered, with the ground state no longer connected to one of the
excited states, but rather
diverging to $+\infty$ along the line $N=1$. Then for $l>0$ 
the ground state eigenvalue ventures for the first time into the
domain $N\in (0,1]$.
(To make sense of the eigenvalue problem in this regime requires its
consideration on a cover of the once-punctured complex plane, with the
sectors $\CS_{-1}$ and $\CS_1$ overlapping 
on different sheets.)  Figures 2e and 2f 
show this behaviour. Observe from figure 2f that
the ground state eigenvalue is zero at 
$l=0.05$, $N=0.2$. 
More generally, (\ref{czform}) implies that
the problem has a zero eigenvalue when $N=4l$.
This means that the steeply-climbing line on figure 2e should also
return to zero, at $N=0.004$. While our program currently breaks down at
such small values of $N$, plots made at intermediate
values of $l$ (between $0.001$ and $0.05$) offer clear support for this
scenario.

For the values of $l$ pictured, the spectrum is entirely 
real for $N\geq 2$, thus generalising the reality property observed
by Bender and Boettcher for $l=0$.

In a subsequent paper \cite{BBN} 
further eigenvalue problems were discussed, based on
Hamiltonians of the form $p^2+x^{2K}(ix)^{\varepsilon}$, with 
the two `quantisation wedges' chosen so as to be centred on the
positive and negative real axes when $\varepsilon=0$.
In these conventions, the earlier problem corresponds to
$K=1$, $\varepsilon=N{-}2$. It is straightforwardly checked that, just
as $T_{1/2}$ or $C^{(1)}$
encodes the spectrum of the problem for $K=1$, so the
functions $T_{K/2}$ or $C^{(K)}$
encode the spectra of the more general problems.
This means that these problems, treated separately in
\cite{BBN}, are in fact intimately linked together, since their spectral
determinants participate in the fusion hierarchy, or T-system,
discussed in \S\ref{monodsec} above.
On the other hand,
the numerical and analytical work in \cite{BBN} immediately gives
us a rather detailed picture of the behaviour of the zeroes of the
general fused T operators. Worth noting in this regard
is the fact that the first
transition to complex eigenvalues for the Schr\"odinger problem, 
that is to complex 
zero positions for the fused T operator, always occurs at
$\varepsilon=0$. In the earlier notation of this paper, this
point corresponds 
to $2M=N=2K$, and for $K>1$, it lies well {\em inside}\ the
semiclassical domain $M>1$, $\beta^2=1/(M{+}1)<1/2$. 

As this paper was being finished, a further article by Bender {\em et
al} appeared \cite{BBetal}. There a form of `classical limit'
($\beta^2\rightarrow 0$) is discussed. More precisely, the limit
is $M\rightarrow\infty$ with $l=0$ and ${\cal E}\equiv E/M^2$ held
fixed. (This differs from the limit treated in appendix~B of
\cite{BLZ2} where $M\rightarrow\infty$ with $l$ and
$E$ is held fixed, and the functions $A_{\pm}$
and $T_j$, $j=O(M)$, are treated. Taking the limit with $E/M^2$
fixed instead of $E$ is appropriate for capturing the behaviour of the
functions $T_j$ with $j$ remaining finite.) For the initial 
problem,
corresponding to $K=1$ and the function $T_{1/2}$ or $C^{(1)}$,
Bender {\em et al} found
\eq
E_k(M)\sim (k+\fract{1}{2})^2M^2~~,\qquad k=0,1\dots,~~~~~
M\rightarrow\infty\,.
\label{newresl}
\en
(Note that in \cite{BBetal} $M$ is instead used 
to label the different eigenproblems, the r\^ole played by $K$ here.)
The result (\ref{newresl}) suggests that
the limiting form for the spectral determinant
$C^{(1)}({\cal E},l)$ at $l{=}0$ is
$2\cos(\pi\sqrt{-{\cal E}})\,$.
(The minus sign appears for the same reason as before, namely the
variable change in going from (\ref{bbprob}) to (\ref{nbbprob}).)
Assuming that the fusion hierarchy (\ref{csys}) holds in this
limit, from it we deduce
\eq
C^{(K)}({\cal E},0)\sim 
\frac{\sin((K{+}1)\pi\sqrt{-{\cal E}})}{\sin(\pi\sqrt{-{\cal E}})}~.
\label{ckgen}
\en
This correctly reproduces the spectra for the higher problems that
in \cite{BBetal} had to be obtained
by a series of independent calculations.
Finally, we remark that the angular momentum term $l(l{+}1)x^{-2}$
drops out of (\ref{gbbprob}) in the limit considered in \cite{BBetal},
and so the results (\ref{newresl}) and (\ref{ckgen}) should hold unchanged
for $l\neq 0$.

\resection{Singular potentials and duality}
\label{dualsec}
Up to now the parameter  $M$ has been restricted to the range 
$(0,\infty)$, but  simple  transformation properties
of the Schr\"odinger equation allow a wider range 
of $M$ to be accommodated. 
The idea is to define a mapping between the eigenvalue problem  
for $M>0$  and the problem with $-1<M<0$.
Our discussion here is not meant to  be 
particularly original, essentially following \cite{BLZschr}, but we
wish to underline the fact that the relevance
of integral equations to Schr\"odinger problems, stressed
in~\cite{DT3}, actually applies to wider range 
of problems than might initially be thought. 
The first step is to make Langer's \cite{langer} variable change
\eq
x=e^{z}~~~,~~~ y(x,E,l)=e^{z/2} \psi(z,E,l)\,.
\en
Then
\eq
\lf(- {d^2 \over dz^2} + e^{2 z(M+1)}  
-E  e^{2 z} \ri) \psi(z,E,l)= \kappa^2 \psi(z,E,l)
\en
with $\kappa=-i (l +\fract{1}{2}) $.
Now we interchange the r\^ole of the two exponential terms, with the 
transformation
\eq
z \rightarrow \fract{1}{M{+}1}\, z   + 
\log\fract {M{+}1} 
{\sqrt{E}~} 
\en
to obtain 
\eq
\lf(-{d^2 \over d z^2}  
 -  e^{2z/(M{+}1)}+\tilde{E}e^{2z} \ri) \psi(z,E,l)= 
\tilde{\kappa}^2 \psi(z,E,l)\,,
\en
where $\tilde{\kappa}=\kappa/(M{+}1)\,$, $\tilde l=(l{-}M/2)/(M{+}1)$ 
and 
\eq
\tilde{E}=\frac{-1~~}{(M{+}1)^{2 M}}\, E^{-M-1}\,.
\label{dualE}
\en
It is easy to check that for $M=1$ (\ref{dualE}) 
gives  an exact mapping  between 
the energy levels of the harmonic oscillator at angular momentum $l$
and those of the Coulomb potential at angular momentum $\tilde l=l/2-1/4$.
Indeed it is straightforward to see 
that  eigenfunctions of the original problem  are transformed into
eigenfunctions  for the `dual' problem: reversing the original variable
change, we have that
\eq
\tilde{y}(x,\tilde{E},\tilde{l})=
(M{+}1)^{-1/2} E^{1/4}
 x^{M \over 2(M+1)} 
y ((M{+}1)E^{-1/2} 
x^{1/(M{+}1)}, E, l)
\en
solves
\eq
\lf( -{d^2 \over d x^2}  -x^{2 \tilde{M}}+\frac{\tilde{l}(\tilde{l}{+}1)}{x^2}
- \tilde{E}  \ri) \tilde{y}(x,\tilde{E},\tilde{l}) 
= 0~,  
\en
with
$ \tilde{M }= (M{+}1)^{-1}  -1 $.
Furthermore, if $y$ decays as $x\rightarrow +\infty$ then so does
$\tilde y$, and
if $y \sim x^{l+1} $ near $x=0$  then
$\tilde{y} \sim x^{\tilde{l}+1}$. 
Hence eigenfunctions are mapped to eigenfunctions,
and the relation (\ref{dualE}) is true at the
level of the eigenvalues too. 
A curious consequence of this transformation, applied
to the problem with $M=1/2$ and $l=0$, is that the solution
of the $x^{-2/3}$ potential with $\tilde{l}= -1/6$ can be 
written in terms of the Airy function. 

The mapping to singular potentials, decaying (albeit algebraically)
at infinity, highlights the fact that the motion of the
zeroes of the ${\bf Q}$-operators as a function of the `twist' $p$
corresponds to the motion of Regge poles for the radial 
Schr\"odinger equation as the angular momentum $l$ is continuously
varied \cite{regge,Euan,NEWT}.
Quite apart from the intrinsic curiosity of this fact, it means
that  the general results of, for example, chapters 8 and 13 of
\cite{NEWT} can be applied
to deduce features of the motion of the zeroes of
the ${\bf Q}$-operators as $p$ and $\beta^2$, or equivalently 
$l$ and $E$, are varied.
Observe also that, from the point of view of
integrable field theory,
the duality $M\rightarrow \tilde M$ sends $\beta^2$ to $1/\beta^2$.
As $M$ passes
through $0$ into the `dual' regime $-1<M<0$, $\beta^2$ increases through
$1$ and the associated perturbation of a $c=1$ conformal field theory
becomes irrelevant.
The full range $-1<M<\infty$ is mapped onto $0<\beta^2<\infty$. 
For
$M<-1$, the nature of the Schr\"odinger problem
(\ref{sibeq}) changes fundamentally, as the singularity at the origin
ceases to be regular. Formally $M<-1$ corresponds to purely imaginary
values of $\beta$, and
it is tempting to suppose that there should be a
relation with the non-compact
sinh-Gordon model. However at this stage this is pure speculation.

%
%%XXXXXXXXXXXXX
\resection{Conclusions}
\label{concsec}
In this paper we have further explored the correspondence between
certain integrable quantum field theories and 
Schr\"odinger equations. The two natural classes
of spectral problems for these equations have been
associated with the two types of operators which arise in the
integrable quantum field theories, namely the ${\bf T}$ and 
${\bf Q}$-operators. The neatness of this fit convinces us that the
matching is no accident, and that it promises to lead to a fruitful
interaction between these previously distinct fields. A new
angle on the theory of integrable models seems to be opening up, and
this alone should justify some further work on the many unresolved
issues which remain.

On a technical level, while we have tried to give most of the details
of our calculations, our discussion has not been
completely rigorous and some aspects will probably repay a more careful
treatment. In addition, the treatment of this paper should be completed
with a full discussion of the `quantum' (non-semiclassical) domain $M<1$.
This is particularly important because the region includes the
unitary minimal models.
One example, the `Airy case' ($M=1/2$, $l=0$), has already been
treated in \cite{V4}, and its consistency with integrable field theory
predictions has been confirmed both numerically~\cite{DT3} and
analytically~\cite{paul}. 

The literature on the spectral theory of the (radial)
Schr\"odinger equation is extensive, much of it dating back
to Regge pole theory, and we can hope to apply further results to the
study of integrable quantum field theories. 
In the other direction, the realisation that the some of the
functional relations uncovered 
by Sibuya, Voros and others in connection with ordinary differential
equations also arise in integrable models, and can be solved by means
of nonlinear integral equations, promises to be very useful. It should
be remarked that functional relations can be found for more general
potentials than the $x^{2M}+l(l{+}1)x^{-2}$ cases used above (see
\cite{Si,V6,V8}). It would be interesting to know if these also have a
r\^ ole to play in the theory of integrable models.
A relation between certain functional relations and differential
equations was recently discussed in \cite{FS}, though the
approach is rather different from that adopted here.

The match that we have found connects spectral determinants of the
Schr\"odinger equation with the vacuum expectation values
of the ${\bf T}$ and ${\bf Q}$-operators.  However 
this leaves to one side all of the other expectation values, which
emerge when excited states in the integrable quantum field theories
are discussed. They obey the same sets of functional relations 
as the vacuum expectation values \cite{KP,BLZ1,BLZ2,BLZ3}, 
and a major challenge is to find an interpretation for them
within the context of Schr\"odinger equations. One way to access 
excited states in integrable quantum field theories is by a process of
analytic continuation in a suitable parameter~\cite{DT1}. However
if we remain with the massless models discussed in this paper, 
this idea is unlikely to help since
the functions $\psi^{\pm}$ and $y$ are 
single-valued in the one remaining
parameter, namely $l$ (see \cite{NEWT,HC}).
If the correspondence could be extended to cover massive models, then
the story would change since these models are already known to
exhibit a pattern of branch points joining the 
ground state to excited states (examples can be found in
\cite{DT1,DT2}).  Whether massive models can be
incorporated into the differential equation framework is another
question that for the moment remains completely open, and any progress
on this issue would be extremely interesting.

\medskip
%%%%%%%%%%%%%%%%%%%%%%%%%%%%%%%%
\noindent {\bf Acknowledgements --} We are grateful to
Armen Allahverdyan,
Clare Dunning, Paul Fendley, Davide Fioravanti,
Ferdinando Gliozzi, Bernard Nienhuis and Paul Pearce
for useful discussions.
The work was supported in part by a TMR grant of the
European Commission, reference ERBFMRXCT960012.
PED  thanks the EPSRC for an Advanced Fellowship, and
RT thanks the Universiteit van Amsterdam for a post-doctoral
fellowship.
%%%%%%%%%%%%%%%%%%%%%%%%%%%%%%%%

\setcounter{section}{0}
\renewcommand{\theequation}{\Alph{section}.\arabic{equation}}

\appsection{WKB details}
In this appendix we sketch how to obtain 
(\ref{Dlim}). The equation
describes the asymptotic behaviour of $\log D^-(E,l)=
W[y(x,E,l),\psi^+(x,E,l)]$. 
Note first that $D^-$ can be evaluated as
\eq
D^-(E,l)=\lim_{x\rightarrow 0}\lf[(2l{+}1)\,x^l\,y(x,E,l)\ri]\,.
\label{daltform}
\en
For convenience we set $E=-a$, so that the problem under
discussion becomes
\eq
\lf(-\frac{d^2}{dx^2}+x^{2M}+\frac{l(l{+}1)}{x^2}+a\ri) y = 0
\en
and we are interested in the behaviour as $|a|\rightarrow +\infty$.
The WKB 
approximation cannot be applied directly to this
equation, because of the nature of its singularity at $x=0$ \cite{langer}.
The remedy found in~\cite{langer} is to make the variable change
$x=e^z$,
$y(x)=e^{z/2}\phi(z)$. Writing $\lambda=l+1/2$ the equation becomes
\eq
\lf(-\frac{d^2}{dz^2}+R(z,a,\lambda)\ri) \phi = 0
\label{tranfeq}
\en
with $R(z,a,\lambda)=e^{(2M{+}2)z}+ae^{2z}+\lambda^2$.
The WKB approximation for $\phi$ is
\eq
\frac{A}{R(z,a,\lambda)^{1/4}}\exp%
\lf[\int_{z_0}^z\!\!\sqrt{R(u,a,\lambda)}\,du\ri]
+ \frac{B}{R(z,a,\lambda)^{1/4}}\exp%
\lf[-\!\int_{z_0}^z\sqrt{R(u,a,\lambda)}\,du\ri]~.
\en
In contrast to the WKB treatment of the initial equation, this is 
approximation is good
for all real $z$ (and hence for $x$ all the way down to zero)
since $R^{-3/4}(R^{-1/4})''$
tends to zero uniformly in $z$ as $|a|\rightarrow +\infty$, so long as
$\arg a\neq\pi$ (for a 
discussion, see~\cite{olver}).
 
The solution that we hope to approximate
is subdominant as $x\rightarrow +\infty$, which requires
$A=0$. The value of $B$ must be chosen so as to match
the large-$x$ asymptotic (\ref{asrep}). A little thought shows 
that the correctly-normalised
approximate solution can be written as
\eq
\phi(z,a,\lambda)\sim \frac{1}{R(z,a,\lambda)^{1/4}}
\exp\lf[\int_{z}^{\infty}\![
\sqrt{R(u,a,\lambda)}-e^{(M{+}1)u}]du
 -\frac{1}{M{+}1}e^{(M{+}1)z}\ri]
\label{normsol}
\en
To estimate (\ref{daltform}) we need to analyse this
quantity as $z\rightarrow -\infty$. An integration by parts extracts a term
$-\lambda z$ in this limit, after which $\lambda$
can be dropped from the remaining integral, giving a 
leading $a$-dependence 
\bea
\phi(z,a,\lambda)&\sim &
\lambda^{-1/2}\exp\lf[-\lambda z
+\int_{-\infty}^{\infty}\![\sqrt{e^{(2M{+}2)u}{+}ae^{2u}}
-e^{(M{+}1)u}]\,du\,\ri]\nn\\[3pt]
&&=
\lambda^{-1/2}\exp\lf[-\lambda z+
a^{(M{+}1)/2M}\!\int_0^{\infty}\![\sqrt{t^{2M}{+}1}-t^M]\,dt
\,\ri]\,.
\label{leading}
\eea
Reverting to the original variables and substituting into
(\ref{daltform}), it is easily checked that the
result quoted in the main text is recovered.

If $M<1$, the above treatment fails, because the integral in
(\ref{normsol}) diverges. The remedy is to introduce further 
regularising terms, by expanding $\sqrt{R(u,a,\lambda)}$ to higher 
order. For example, if $1/3<M<1$ the argument of the exponential in
(\ref{normsol}) should
be replaced by
\eq
\lf[\int_{z}^{\infty}\![
\sqrt{R(u,a,\lambda)}-e^{(M{+}1)u}-\frac{a}{2}e^{(1{-}M)u}]\,du
 -\frac{1}{M{+}1}e^{(M{+}1)z}-
 \frac{a}{2{-}2M}e^{(1{-}M)z}\ri]
\en
and (\ref{leading}) becomes
\eq
\phi(z,a,\lambda)\sim 
\lambda^{-1/2}\exp\lf[-\lambda z+
a^{(M{+}1)/2M}\!\int_0^{\infty}\![\sqrt{t^{2M}{+}1}
-t^M-\fract{1}{2}t^{-M}]\,dt
\,\ri]\,.
\en
However, for $1/3<M<1$ we have
\eq
2 \int_0^{\infty}[(t^{2M}+1)^{1/2}-t^M
-\fract{1}{2}t^{-M}]\,dt
=-\frac{1}{\sqrt\pi}
\Gamma(-\fract{1}{2}-\fract{1}{2M})\Gamma(1+\fract{1}{2M})\,,
\en
and so the 
result (\ref{intresult}) is preserved -- we have simply continued
it around the pole at $M=1$. Despite the fact that the $|E|\rightarrow
\infty$ asymptotic of
$D^-$ is unchanged, the move to $M<1$ does
have an effect  on the limiting behaviour of
the auxiliary function $d(E,l)\equiv
\omega^{2l+1}{D^-(\omega^2E,l)}/{D^-(\omega^{-2}E,l)}$ used in
\S\ref{tqsec}. The correct choice of branches when applying
the asymptotic (\ref{Dlim}) at the shifted arguments 
$\omega^{\pm 2\!}E$ changes, with the result that 
the asymptotic
$\log d(E,l)\sim -\fract{1}{2}ib_0 (E)^{\mu} $ now holds in the
revised
sector $-\frac{2\pi M}{M{+}1}<\arg (E)<\frac{2\pi M}{M{+}1}\,$.
Outside this sector, the leading contributions cancel and so all
that can be concluded is that the asymptotic is subleading, though from
the nonlinear integral equation (\ref{nlie})
we expect that it will be $O(1)$, and 
in fact equal to $(2l{+}1)\pi i$.

\end{document}